\documentclass[%
reprint,
superscriptaddress,
amsmath,
amssymb,
aps,
pra,
]{revtex4-2}

\usepackage{graphicx}% Include figure files
\graphicspath{{./},{figs/},}

\usepackage{dcolumn}% Align table columns on decimal point
\usepackage{bm}% bold math
\usepackage{physics}
\usepackage{algorithm,algpseudocode,tabularx,multirow}

\usepackage[normalem]{ulem}

\usepackage[colorlinks=true,citecolor=blue,linkcolor=magenta]{hyperref}
\usepackage{cleveref}
    \crefname{subsection}{Subsection}{Subsections}

\usepackage[x11names]{xcolor}

\usepackage{adjustbox}
\usepackage{tikz}
\usetikzlibrary{quantikz2}

\makeatletter
\let\old@makecaption=\@makecaption
\usepackage{subcaption}
\let\@makecaption=\old@makecaption
\makeatother

\usepackage{listings}

\begin{document}

\preprint{APS/123-QED}

\title{Fault-tolerant correction-ready encoding of the [[7,1,3]] Steane code on a 2D grid}

\author{Andrea Rodriguez-Blanco}
    \thanks{These authors contributed equally}
    \affiliation{Department of Chemistry, University of California, Berkeley, CA 94720, USA}

\author{Ho Nam Nguyen}
    \thanks{These authors contributed equally}
    \affiliation{Department of Physics, University of California, Berkeley, CA 94720, USA}
    \email{honamnguyen@berkeley.edu \\ rodriguezblanco@berkeley.edu}

\author{K.~Birgitta Whaley}
\affiliation{Department of Chemistry, University of California, Berkeley, CA 94720, USA}

\date{\today}% It is always \today, today,
             %  but any date may be explicitly specified

\begin{abstract}
Practical quantum computation heavily relies on the ability to perform quantum error correction in a fault-tolerant manner. Fault-tolerant encoding is a critical first step, and careful consideration of the error correction cycle that follows is essential for ensuring the encoding's effectiveness and compatibility. In this work, we investigate various correction-ready encoding methods to fault-tolerantly prepare the $\ket{0}_{\rm L}$ state of the [[7,1,3]] Steane code on a 2D grid. Through numerical simulations, we demonstrate that parity-check encoding with a few \textit{Flag-Bridge} qubits outperforms verification-based encoding by achieving lower error rates and allowing flexible tuning of the performance-efficiency trade-off. Additionally, parity-check approach enables a compact \textit{hybrid} protocol that combines encoding and error correction, capable of matching the performance of a standalone error correction protocol with perfect encoding. Surprisingly, compared to the resource-intensive Steane error correction, this low-overhead method still offers a practical advantage in noisy settings. These findings highlight the approach with \textit{Flag-Bridge} qubits as a robust and adaptable solution for noisy near-term quantum hardware.
\end{abstract}

%\keywords{Suggested keywords}%Use showkeys class option if keyword
                              %display desired
\maketitle
%\tableofcontents

\section{Introduction}\label{sec:intro}

To fully harness the power of quantum computing for complex scientific and industrial applications, quantum processors must scale efficiently while adhering to fault-tolerant designs. However, qubits are highly susceptible to errors due to environmental interactions and imperfect quantum gate operations, making the management of error essential for reliable computation. The theory of fault-tolerant quantum computation~\cite{shor1997faulttolerantquantumcomputation,aharonov1999faulttolerantquantumcomputationconstant,Kitaev_1997,Knill1998-ng,gottesman1997stabilizercodesquantumerror} provides the necessary tools to mitigate these errors and enable scalability. 

Within this framework, quantum algorithms are not executed directly on physical qubits but instead on logical qubits encoded using a quantum error-correcting (QEC) code \cite{Nielsen_Chuang_2010}. The purpose of a QEC code is to protect quantum information by redundantly encoding it within a subspace of a larger physical system, known as the \textit{codespace}. During computation, errors on physical qubits may displace logical states from this subspace, but by interleaving error correction throughout the process, the logical states can be effectively restored. Consequently, a properly designed QEC code can tolerate some level of physical errors while preserving logical information. The threshold theorem \cite{Knill1998-ng,thresholds_memory,gottesman1997stabilizercodesquantumerror} formalizes this principle, guaranteeing that if the physical error rate $p$ remains below a critical threshold $p_{\text{th}}$, logical error rates $p_{\rm L}$ can be exponentially suppressed, enabling arbitrarily long quantum computations.

In practice, a QEC protocol -- consisting of encoding, error detection, and correction -- is implemented using imperfect gates and measurements. Ensuring that each component is fault-tolerant introduces additional complexity \cite{DiVincenzoShor, DiVincenzoAliferis, Steane1996_FTQEC, Knill2005, aliferis2005quantumaccuracythresholdconcatenated, Yoder2017surfacecodetwist, ChaoReichardt2018_twoextraqubits}. For example, Steane-style parity checks double the qubit overhead~\cite{Steane1996_FTQEC}, while flag-based techniques require fewer extra qubits but involve more gate operations and longer circuits to detect faults~\cite{Yoder2017surfacecodetwist, ChaoReichardt2018_twoextraqubits}. A realistic evaluation of a QEC code must account for the actual implementation of these components using noisy gates to determine how effectively errors are suppressed at the logical level. This assessment can be performed through simulations under specific noise models or validated in physical experiments.

Recent advancements in QEC have been driven by the constraints of near-term quantum hardware, including limited physical qubits and architectural connectivity. Significant milestones have been achieved, such as the reduction of logical error rates for memory qubits as surface code distance increases \cite{Google2023, Google2024}, the development of quantum low-density parity-check (qLDPC) codes to maintain constant encoding rates while mitigating non-locality overheads \cite{AndrewLucas2024_4logicals, Gottesman2025_2DqLDPC}, and the use of novel concatenated codes to further reduce logical error rates while improving encoding efficiency \cite{paetznick2024demonstrationlogicalqubitsrepeated, Yamasaki2024, yoshida2024concatenatecodessavequbits, Goto2024hypecubes}.

Among these developments, the [[7,1,3]] Steane code \cite{Steane1996_FTQEC}, the smallest instance of color codes \cite{colorcodesbombin}, remains a key testbed for fault-tolerant quantum computation. In recent years, it has been instrumental in demonstrating critical fault-tolerant capabilities, including real-time error correction \cite{RyanAnderson2021} and the implementation of a universal logical gate set, such as transversal Clifford gates \cite{ryananderson2022implementingfaulttolerantentanglinggates, Quantiniuum2024_teleportinglogicalqubit, Bluvstein2024, Postler2022} and non-Clifford gates \cite{Postler2022, rodriguez2024experimentaldemonstrationlogicalmagic}. With fault-tolerant protocols successfully validated using the [[7,1,3]] Steane code, efforts to scale toward large-scale universal quantum computation are now focused on increasing the distance of color codes \cite{ryananderson2022implementingfaulttolerantentanglinggates, rodriguez2024experimentaldemonstrationlogicalmagic} and employing concatenation with error-detecting codes or other quantum Hamming codes \cite{yoshida2024concatenatecodessavequbits, Yamasaki2024}.

% However, so far, all the experimental realizations of the [[7,1,3]] Steane code have been based on architectures with all-to-all connectivity  In this paper, we focus on studying the 2D local implementation of various fault-tolerant encoding and error-correcting protocols for the Steane code. The motivation is twofold: first, to encourage the experimental implementation of the Steane code on platforms with connectivity constraints, such as superconducting qubits; and second, to investigate schemes that minimize movement, thereby reducing decoherence effects caused by ion shuttling and the reconfiguration of atomic arrays that can become more relevant for scalable quantum processors. In the near term, we will still face limitations in the number of logical qubits available on a platform and in logical error rates, which set a constraint on the depth of quantum circuits. Therefore, optimizing the performance of the [[7,1,3]] code remains a topic of worthwhile interest.

Thus far, all experimental realizations of the [[7,1,3]] Steane code have relied on architectures with all-to-all connectivity. In this work, we focus on studying its implementation in a 2D grid architecture with only local interactions, exploring fault-tolerant encoding and error-correction protocols that are adapted to realistic connectivity constraints. Our motivation is twofold: first, to encourage experimental realizations of the Steane code on hardware platforms with limited connectivity, such as superconducting qubits; and second, to develop schemes that minimize qubit movement, thereby reducing decoherence effects caused by ion shuttling \cite{Ruster2014,Moses2023_racetrack, ovide2024scalingassigningresourcesion,loschnauer2024scalablehighfidelityallelectroniccontrol} and atomic array reconfiguration \cite{Bluvstein2024}---challenges that become increasingly significant as quantum processors are scaled up. In the near term, quantum hardware will continue to be constrained by the limited availability of logical qubits and non-negligible logical error rates, which impose restrictions on the depth of quantum circuits. As a result, optimizing the performance of the [[7,1,3]] Steane code remains an important and timely pursuit.

Several theoretical studies have explored aspects of implementing the [[7,1,3]] Steane code in a 2D architecture~\cite{LaoAlmudever2020, zen2024quantumcircuitdiscoveryfaulttolerant}. Ref.~\cite{LaoAlmudever2020} embeds the [[7,1,3]] code using \textit{Flag-Bridge} qubits to address connectivity constraints while preserving the fault-tolerant nature of parity-check circuits. However, that study assumes a standalone error correction protocol with perfect encoding, overlooking imperfections in state preparation that could degrade performance. Conversely, Ref.~\cite{zen2024quantumcircuitdiscoveryfaulttolerant} investigates the embedding of a verification-based fault-tolerant encoding circuit~\cite{Goto2016} for the [[7,1,3]] code into a 2D grid. A reinforcement learning agent is used to efficiently map the standard encoding circuit onto a square lattice, optimizing qubit resources and reducing circuit depth. However, that work does not analyze the error-correction (EC) cycle following the encoding, potentially underestimating the benefits of optimized circuits when integrated into a full fault-tolerant QEC protocol.

In the current work, we investigate the complete stack for  near term fault-tolerant implementation the [[7,1,3]] Steane code, integrating various logical state preparation circuits and error-correction schemes within a square lattice with only nearest-neighbor interactions. We design multiple fault-tolerant encoding circuits that are either inherently compatible with a given error-correction method, or require only minimal reconfiguration. Furthermore, we leverage overlapping components between encoding and error-correction gadgets to develop an optimized \textit{hybrid} protocol that preserves overall level-1 fault tolerance while efficiently managing logical errors.

The paper is structured as follows. In Sec.~\ref{Sec:background}, we review the properties of the [[7,1,3]] Steane code and fault-tolerant error-correction protocols based on flag qubits and the Steane method. Sec.~\ref{Sec:enc} introduces two encoding protocols: one utilizing fault-tolerant parity-checks with flag qubits and the other employing verification qubits. In Sec.~\ref{sec:numerical}, we present simulation results for different combinations of these encoding and error-correction strategies. We first analyze correction-ready encoding circuits in isolation before evaluating various approaches in integrating encoding and error-correction. Finally, in Sec.~\ref{Sec:conclusions}, we summarize our findings and outline potential future directions.

\section{Background}\label{Sec:background}
We begin with a brief overview of stabilizer codes~\cite{gottesman1997stabilizercodesquantumerror}, a class of QEC codes to which the [[7,1,3]] Steane code belongs. Stabilizer codes are defined by an abelian subgroup $\mathcal{S}$ of the $n$-qubit Pauli group $\mathcal{P}_{n}$. The stabilizer group $\mathcal{S}$ is generated by $r$ independent stabilizer generators  (or commonly referred to as \emph{stabilizers}) $\{S_1,S_2, ..., S_r\}$, where each stabilizer $S_i\in P_n$ satisfies $S_i^2=I$ and commutes with all other elements in $\mathcal{S}$. The code subspace is defined as the simultaneous $+1$-eigenspace of all stabilizers, meaning any valid codeword, or logical state, $\ket{\psi}$ satisfies $ S_i\ket{\psi} = \ket{\psi}$ for all $S_i$. All other eigenspaces correspond to error subspaces with eigenvalue $-1$ for at least one stabilizer.

Each stabilizer imposes one constraint on the original $2^n$-dimensional Hilbert space of $n$ physical qubits, leaving  $2^{n-r}$ degrees of freedom to encode $k=n-r$ logical qubits in the code space. When an error $E$ occurs, measuring the stabilizers projects the corrupted logical state into either the code subspace, if $E$ commutes with all stabilizers, or one of the error subspace otherwise. The resulting measurement outcomes, known as the \emph{syndrome}, provides crucial information to detect whether the logical state has left the code subspace, and if so, to determine the necessary recovery operation $R$. The maximum weight of a correctable error is determined by the code distance, defined as the smallest number of qubits that must be altered for one valid codeword to transform into another. Specifically, a code with distance $d$ can correct errors with weight up to $\lfloor(d-1)/2)\rfloor$. Altogether, the parameters $[[n,k,d]]$ capture the key properties of a QEC code.

\subsection{[[7,1,3]] Steane code}\label{subsec:713_Steane_code}

\begin{table}[t!]
\begin{center}
    \begin{tabular}{|c|c|} 
        \hline Syndrome $\left(s^X/s^Z\right)$& Error (Recovery) \rule{0pt}{2.5ex}\\
        \hline \hline
        000 & $I$ \\ 
        100 & $Z_1/X_1$ \\ 
        110 & $Z_2/X_2$ \\ 
        111 & $Z_3/X_3$ \\ 
        101 & $Z_4/X_4$ \\ 
        010 & $Z_5/X_5$ \\ 
        011 & $Z_6/X_6$ \\ 
        001 & $Z_7/X_7$ \\ 
        \hline
    \end{tabular}
    \caption{Standard look-up-table for the [[7,1,3]] Steane code, mapping each syndrome to the corresponding error that needs to be corrected. Each syndrome corresponds to measurement outcomes of the stabilizers $\{S_1^{X/Z},S_2^{X/Z},S_3^{X/Z}\}$, and the recovery operation is simply the application of the identified error to restore the original codeword.}
    \label{tab:s2LUT}
\end{center}
\end{table}

% \begin{figure*}[t!]
%     \centering
%     \begin{subfigure}{0.35\textwidth}
%     \centering
%         \begin{quantikz}[row sep=0.1cm,column sep=0.1cm]
%             \qw & \lstick[4]{data} & \targ{}\vqw{0} &\qw &\qw &\qw &\qw &\qw &\qw &\qw\\
%             \qw &  &\qw &\qw & \targ{}\vqw{0} &\qw &\qw &\qw &\qw &\qw\\
%             \qw &  &\qw &\qw &\qw & \targ{}\vqw{0} &\qw &\qw &\qw &\qw\\
%             \qw &  &\qw &\qw &\qw &\qw &\qw & \targ{}\vqw{0} &\qw &\qw\\
%             \qw & \lstick[1]{syndrome~~$\ket{+}$} & \ctrl{-4} & \ctrl{0} & \ctrl{-3} & \ctrl{-2} & \ctrl{0} & \ctrl{-1} & \meterD{X} & \\
%             \qw & \lstick[1]{flag~~$\ket{0}$} &\qw & \targ{}\vqw{-1} &\qw &\qw & \targ{}\vqw{-1} &\qw & \meterD{Z} &  
%         \end{quantikz}           
%     \end{subfigure}
%     % \hfill
%     \begin{subfigure}{0.35\textwidth}
%     \centering
%         \begin{quantikz}[row sep=0.1cm,column sep=0.1cm]
%             \qw & \lstick[4]{data} & \targ{}\vqw{0} &\qw &\qw &\qw &\qw &\qw &\qw &\qw\\
%             \qw &  &\qw &\qw & \targ{}\vqw{0} &\qw &\qw &\qw &\qw &\qw\\
%             \qw &  &\qw &\qw &\qw & \targ{}\vqw{0} &\qw &\qw &\qw &\qw\\
%             \qw &  &\qw &\qw &\qw &\qw &\qw & \targ{}\vqw{0} &\qw &\qw\\
%             \qw & \lstick[1]{syndrome~~$\ket{+}$} & \ctrl{-4} & \ctrl{0} & \ctrl{-3} & \ctrl{-2} & \ctrl{0} & \ctrl{-1} & \meterD{X} & \\
%             \qw & \lstick[1]{flag~~$\ket{0}$} &\qw & \targ{}\vqw{-1} &\qw &\qw & \targ{}\vqw{-1} &\qw & \meterD{Z} &  
%         \end{quantikz}        
%     \end{subfigure}
% \caption{Test figure}
% \end{figure*}

\begin{figure*}[t!]
\centering
\begin{subfigure}{0.6\textwidth}
    \centering 
    \begin{quantikz}[row sep=0.15cm,column sep=0.1cm]
        \lstick[7]{data} &\qw &\qw \gategroup[14,steps=9,style={dashed,rounded corners,fill=blue!20, inner xsep=1pt},background]{$X$-check} & \targ{}\vqw{0} &\qw &\qw &\qw &\qw &\qw &\qw &\qw &\qw &\qw &\qw \gategroup[14,steps=9,style={dashed,rounded corners,fill=red!20, inner xsep=1pt},background]{$Z$-check} & \ctrl{0} &\qw &\qw &\qw &\qw &\qw &\qw &\qw &\qw\\
        &\qw &\qw &\qw & \targ{}\vqw{0} &\qw &\qw &\qw &\qw &\qw &\qw &\qw &\qw &\qw &\qw & \ctrl{0} &\qw &\qw &\qw &\qw &\qw &\qw &\qw\\
        &\qw &\qw &\qw &\qw & \targ{}\vqw{0} &\qw &\qw &\qw &\qw &\qw &\qw &\qw &\qw &\qw &\qw & \ctrl{0} &\qw &\qw &\qw &\qw &\qw &\qw\\
        &\qw &\qw &\qw &\qw &\qw & \targ{}\vqw{0} &\qw &\qw &\qw &\qw &\qw &\qw &\qw &\qw &\qw &\qw & \ctrl{0} &\qw &\qw &\qw &\qw &\qw\\
        &\qw &\qw &\qw &\qw &\qw &\qw & \targ{}\vqw{0} &\qw &\qw &\qw &\qw &\qw &\qw &\qw &\qw &\qw &\qw & \ctrl{0} &\qw &\qw &\qw &\qw\\
        &\qw &\qw &\qw &\qw &\qw &\qw &\qw & \targ{}\vqw{0} &\qw &\qw &\qw &\qw &\qw &\qw &\qw &\qw &\qw &\qw & \ctrl{0} &\qw &\qw &\qw\\
        &\qw &\qw &\qw &\qw &\qw &\qw &\qw &\qw & \targ{}\vqw{0} &\qw &\qw &\qw &\qw &\qw &\qw &\qw &\qw &\qw &\qw & \ctrl{0} &\qw &\qw\\
        \lstick[7]{ancilla} &\qw & \gate[7]{\ket{0}_{\rm L}} & \ctrl{-7} &\qw &\qw &\qw &\qw &\qw &\qw & \meterD{X} &\qw &\qw & \gate[7]{\ket{+}_{\rm L}} & \targ{}\vqw{-7} &\qw &\qw &\qw &\qw &\qw &\qw & \meterD{Z} & \\
        &\qw &\qw &\qw & \ctrl{-7} &\qw &\qw &\qw &\qw &\qw & \meterD{X} &\qw &\qw &\qw &\qw & \targ{}\vqw{-7} &\qw &\qw &\qw &\qw &\qw & \meterD{Z} & \\
        &\qw &\qw &\qw &\qw & \ctrl{-7} &\qw &\qw &\qw &\qw & \meterD{X} &\qw &\qw &\qw &\qw &\qw & \targ{}\vqw{-7} &\qw &\qw &\qw &\qw & \meterD{Z} & \\
        &\qw &\qw &\qw &\qw &\qw & \ctrl{-7} &\qw &\qw &\qw & \meterD{X} &\qw &\qw &\qw &\qw &\qw &\qw & \targ{}\vqw{-7} &\qw &\qw &\qw & \meterD{Z} & \\
        &\qw &\qw &\qw &\qw &\qw &\qw & \ctrl{-7} &\qw &\qw & \meterD{X} &\qw &\qw &\qw &\qw &\qw &\qw &\qw & \targ{}\vqw{-7} &\qw &\qw & \meterD{Z} & \\
        &\qw &\qw &\qw &\qw &\qw &\qw &\qw & \ctrl{-7} &\qw & \meterD{X} &\qw &\qw &\qw &\qw &\qw &\qw &\qw &\qw & \targ{}\vqw{-7} &\qw & \meterD{Z} & \\
        &\qw &\qw &\qw &\qw &\qw &\qw &\qw &\qw & \ctrl{-7} & \meterD{X} &\qw &\qw &\qw &\qw &\qw &\qw &\qw &\qw &\qw & \targ{}\vqw{-7} & \meterD{Z} & 
    \end{quantikz}
    \caption{Steane error-correction}
\end{subfigure}
\hfill
\begin{subfigure}{0.38\textwidth}
\centering
    \begin{minipage}{\textwidth}
    \centering
    \begin{subfigure}{\textwidth}
        \centering
        \begin{quantikz}[row sep=0.02cm,column sep=0.1cm]
            \qw & \lstick[4]{data} &&\qw \gategroup[6,steps=14,style={dashed,rounded corners,fill=green!20, inner xsep=2pt},background]{Circuit 1} & \gate[style={starburst,scale=0.5}]{\color{blue}{Z}} &\qw&\qw&\qw&  \targ{}\vqw{0} &\qw &\qw &\qw &\qw &\qw &\qw &\qw &\qw\\
            \qw &  &&\qw&\qw&\qw&\qw&\qw& \qw &\qw & \targ{}\vqw{0} &\qw &\qw &\qw &\qw &\qw &\qw\\
            \qw &  &&\qw&\qw&\qw&\qw&\qw& \qw &\qw &\qw &\qw & \targ{}\vqw{0} &\qw &\qw &\push{\color{red}{X}} &\qw\\
            \qw &  &&\qw&\qw&\qw&\qw&\qw& \qw &\qw &\qw &\qw &\qw &\qw & \targ{}\vqw{0} &\push{\color{red}{X}} &\qw\\
            \qw &&&& \lstick[1]{synd~~$\ket{+}$} &\qw &\qw &\qw& \ctrl{-4} & \ctrl{0} & \ctrl{-3} & \gate[style={starburst,scale=0.6,fill=yellow}]{\color{red}{X}} & \ctrl{-2} & \ctrl{0} & \ctrl{-1} & \push{\color{blue}{Z}} & \meterD{X} & \\
            \qw &&&& \lstick[1]{flag~~$\ket{0}$} &\qw &\qw &\qw &\qw & \targ{}\vqw{-1} &\qw &\qw &\qw & \targ{}\vqw{-1} &\qw &\push{\color{red}{X}} & \meterD{Z} &  
        \end{quantikz}        
        % \caption{Steane EC circuit}
        % \label{fig:three sin x}
    \end{subfigure}
    \begin{subfigure}{\textwidth}
        \centering
        \begin{quantikz}[row sep=0.02cm,column sep=0.1cm]
            \qw & \lstick[4]{data} &&\qw\gategroup[6,steps=13,style={dashed,rounded corners,fill=green!20, inner xsep=2pt},background]{Circuit 2}&\qw&\qw&\qw&\qw&\qw&\qw &\qw & \targ{}\vqw{0} &\qw &\qw &\qw &\qw &\qw\\
            \qw &  &&\qw&\qw&\qw&\qw&\qw&\qw&\qw &\qw &\qw &\qw & \targ{}\vqw{0} &\qw &\qw &\qw\\
            \qw &  &&\qw&\qw&\qw&\qw&\qw&\qw&\qw & \targ{}\vqw{0} &\qw &\qw &\qw &\qw &\qw &\qw\\
            \qw &  &&\qw&\qw&\qw&\qw&\qw&\qw&\qw &\qw &\qw & \targ{}\vqw{0} &\qw &\qw &\qw &\qw\\
            \qw &&&&&&&& \lstick[1]{synd~~$\ket{+}$} & \ctrl{0} &\qw & \ctrl{-4} &\qw & \ctrl{-3} & \ctrl{0} & \meterD{X} & \\
            \qw &&&&&&&& \lstick[1]{flag~~$\ket{0}$} & \targ{}\vqw{-1} & \ctrl{-3} &\qw & \ctrl{-2} &\qw & \targ{}\vqw{-1} & \meterD{Z} & 
        \end{quantikz}        
        % \caption{Steane EC circuit}
        % \label{fig:three sin x}
    \end{subfigure}
    \begin{subfigure}{\textwidth}
        \centering
        \begin{quantikz}[row sep=0.02cm,column sep=0.1cm]
            \qw & \lstick[4]{data} &&\qw\gategroup[7,steps=14,style={dashed,rounded corners,fill=green!20, inner xsep=2pt},background]{Circuit 3}&\qw&\qw&\qw&\qw&\qw&\qw & \targ{}\vqw{0} &\qw &\qw &\qw &\qw &\qw &\qw &\qw\\
            \qw &  &&\qw&\qw&\qw&\qw&\qw&\qw&\qw &\qw & \targ{}\vqw{0} &\qw &\qw &\qw &\qw &\qw &\qw\\
            \qw &  &&\qw&\qw&\qw&\qw&\qw&\qw&\qw &\qw &\qw & \targ{}\vqw{0} &\qw &\qw &\qw &\qw &\qw\\
            \qw &  &&\qw&\qw&\qw&\qw&\qw&\qw&\qw &\qw &\qw &\qw & \targ{}\vqw{0} &\qw &\qw &\qw &\qw\\
            \qw &&&&&&&& \lstick[1]{synd~~$\ket{+}$} & \ctrl{0} & \ctrl{-4} & \ctrl{-3} &\qw &\qw & \ctrl{0} &\qw & \meterD{X} & \\
            \qw &&&&&&&& \lstick[1]{flag 1~~$\ket{0}$} & \targ{}\vqw{-1} & \ctrl{0} &\qw & \ctrl{-3} &\qw & \targ{}\vqw{-1} & \ctrl{0} & \meterD{Z} & \\
            \qw &&&&&&&& \lstick[1]{flag 2~~$\ket{0}$} &\qw & \targ{}\vqw{-1} &\qw &\qw & \ctrl{-3} &\qw & \targ{}\vqw{-1} & \meterD{Z} & 
        \end{quantikz}        
        % \caption{Steane EC circuit}
        % \label{fig:three sin x}
    \end{subfigure}
    \end{minipage}
    \caption{Flag-based error-correction}
\end{subfigure}
\caption{\textbf{Fault-tolerant syndrome extraction circuits}. a) Steane EC consists of two parity-check circuits where syndromes of the same type (either $X$ or $Z$) are extracted simultaneously. Fault tolerance is achieved using transversal CNOT gates, which interact with only one data qubit at a time. The protocol also requires the fault-tolerant preparation of two logical ancilla states, $\ket{0}_{\rm L}$ and $\ket{+}_{\rm L}$. b) In FB EC, flag qubits are used to detect and prevent harmful faults on the syndrome qubits from propagating to multiple data qubits. Circuit 1 illustrates a weight-4 $X$-check that employs CNOT gates to propagate $Z$ errors (blue) from data qubits to the syndrome qubit. It utilizes one flag qubit to catch harmful $X$ errors originating from the middle two CNOT gates. Circuits 2 and 3 demonstrate different strategies for leveraging flag qubits as bridges, effectively reducing the connectivity requirements of the circuit.}
\label{fig:syndrome_extraction_circuits}
\end{figure*}

The [[7,1,3]] Steane code is a QEC code that encodes a single logical qubit into seven physical qubits, capable of correcting any single-qubit error. It is derived from the classical [[7,4,3]] Hamming code via the CSS (Calderbank-Shor-Steane) construction~\cite{Nielsen_Chuang_2010}. Specifically, the Hamming code's parity-check matrix 
\begin{eqnarray}
    &H_{\rm Hamming} = 
    \begin{pmatrix}
    1 & 1 & 1 & 1 & 0 & 0 & 0\\
    0 & 1 & 1 & 0 & 1 & 1 & 0 \\
    0 & 0 & 1 & 1 & 0 & 1 & 1
    \end{pmatrix} = H_X=H_Z,
    \label{eq:parity_check_matrix}
\end{eqnarray}
is used to construct three $X$-type and three $Z$-type stabilizers independently. The resulting stabilizers for the code are given as:
\begin{equation}
    \begin{split}
    & S_1^X=X_1X_2X_3X_4,  \hspace{0.5cm } S_1^Z=Z_1Z_2Z_3Z_4,\\
    & S_2^X=X_2X_3X_5X_6, \hspace{0.5cm } S_2^Z=Z_2Z_3Z_5Z_6,\\
    & S_3^X=X_3X_4X_6X_7, \hspace{0.5cm }  S_3^Z=Z_3Z_4Z_6Z_7.\\
    \end{split}
\end{equation}
% This separation allows the code to independently correct single-qubit $X$ and $Z$ errors.

The six stabilizers define a code space to encode a single logical qubit with basis states $\ket{0}_{\rm L}$ and $\ket{1}_{\rm L}$. The $\ket{0}_{\rm L}$ logical state is a uniform superposition of all 7-qubit codewords satisfying the stabilizer constraints:
\begin{eqnarray}
    \ket{0}_{\rm L} = \frac{1}{\sqrt{8}}\Big( \ket{0000000} &+ \ket{1111000} &+ \ket{0110110} + \nonumber \\
    \ket{0011011} &+  \ket{1001110} &+  \ket{1010101} +  \nonumber\\
   \ket{0101101}  &+ \ket{1100011}& \Big). \label{eq:logical0}
\end{eqnarray}
The $\ket{1}_{\rm L}$ logical state is obtained via application the logical operator $\tilde{X}_{\rm L}=X_1X_2X_3X_4X_5X_6X_7$ as follows:
\begin{eqnarray}
    \ket{1}_{\rm L} &=& \tilde{X}_{\rm L}\ket{0}_{\rm L} \nonumber \\
    &=& \frac{1}{\sqrt{8}}\Big( \ket{111111} + \ket{0000111} \ket{1001001} + \nonumber \\
    &&+\ket{1100100} +  \ket{0110001} +  \ket{0101010}  \nonumber\\
   &&+\ket{1010010}  + \ket{0011100} \Big). \label{eq:logical1}
\end{eqnarray}
For the [[7,1,3]] code, the logical operators can also be expressed in their stabilizer-equivalent form. For convenience, we adopt the following minimum-weight representation of the logical operators for the rest of the paper:
\begin{eqnarray}
    X_{\rm L} = S_2^X\cdot X_1X_2X_3X_4X_5X_6X_7 &= X_1X_4X_7, \nonumber\\
    Z_{\rm L} = S_2^Z\cdot Z_1Z_2Z_3Z_4Z_5Z_6Z_7 &= Z_1Z_4Z_7.
\end{eqnarray}
Note that this minimum weight corresponds precisely to the code distance $d=3$, which represents the smallest number of physical qubit errors required to cause a logical error. Since these operators commute with the stabilizers, they enable logical operations on the encoded qubit without leaving the code space, rendering them undetectable by the stabilizer measurements. 

% \arb{The [[7,1,3]] code can detect and  correct all possible single-qubit Pauli errors, as it can independently handle both $X$-Pauli and $Z$- Pauli errors. } 
% The decoding process for the [[7,1,3]] \arb{can be done with a standard look-up-table.} By measuring each stabilizer, a binary outcome, \arb{referred to as the syndrome,} is obtained. \arb{$Z$-type stabilizers detect $X$-Pauli errors, and $X$-type stabilizers do the same for the $Z$-Pauli errors .} With three $X$-type and three $Z$-type stabilizers, there are $2^3=8$ possible syndromes per type, which correspond to \arb{\sout{7}} \arb{seven} distinct single-qubit errors on the \arb{\sout{7}} \arb{seven} physical qubits, plus the identity (no error).\arb{\sout{ Importantly, $Z$-type stabilizers detect $X$ errors, and vice versa.}} The mapping between syndromes and errors is summarized in Table.~\ref{tab:s2LUT}.

For the single-qubit errors that the [[7,1,3]] code is designed to handle, the error detection and correction process is straightforward. Stabilizer measurements produce a binary string known as the \textit{syndrome}, which uniquely identifies the error.  Thanks to the CSS structure, the code detects each type of single-qubit error independently, $X$-type stabilizers detect $Z$ errors, and $Z$-type stabilizers detect $X$ errors. Each set of three stabilizers generates $2^3 = 8$ possible syndromes, corresponding to the seven single-qubit errors on the physical qubits plus the no-error case. The mapping between syndromes and errors generates a look-up table, summarized in Table~\ref{tab:s2LUT}. Applying a recovery operation equivalent to the identified error then restores the original logical state.

The primary objective of a QEC code is to repeatedly extract syndromes to detect and correct errors. To reliably achieve this goal, stabilizer measurements must be designed to be fault-tolerant (FT), i.e, they should not introduce more errors than the code can handle~\cite{shor1997faulttolerantquantumcomputation, aharonov1999faulttolerantquantumcomputationconstant, OskinChuang}. We shall consider level-1 FT QEC circuits for the [[7,1,3]] code that are built from qubits encoded at level-0, corresponding to physical qubits. The construction of these level-1 circuits must guarantee  that i) a level-1 QEC circuit with no fault takes an input with at most one error to an output with no error, and ii) a level-1 QEC circuit with one fault  takes an input with no errors to an output with at most one error~\cite{aliferis2005quantumaccuracythresholdconcatenated}. From now on, we shall simply refer to level-1 FT as the FT condition.
In the following, we outline two FT error-correction protocols for the [[7,1,3]] code: 
one protocol proposed by Steane~\cite{Steane1996_FTQEC}, together with a more recently developed protocol that minimizes qubit overhead by utilizing flag qubits~\cite{Yoder2017surfacecodetwist, ChaoReichardt2018_twoextraqubits, ChamberlandBeverland2018_flagFTEC, LaoAlmudever2020}.

\subsection{Steane error-correction}\label{subsec:SteaneEC}

% \arb{In the late 1990s, Steane~\cite{Steane1996_FTQEC} presented an error-correction (EC) protocol more paralleizable than the Shor-style EC method [citeShor]. The Steane's method, although applicable only to CSS codes, prevents the propagation of a single error into multiple errors by utilizing just $2n$ CNOT gates between the data qubits and the ancilla qubits in a two-step syndrome extraction process. Specifically, $n$ gates are used for $X$-type syndrome extraction, while an additional $n$ gates are dedicated to $Z$-type extraction. Syndromes of the same type (either $X$ or $Z$) can be extracted simultaneously in each step (see Table~\ref{tab:s2LUT}).
% For each step, it requires preparing an ancilla state encoded with the same CSS code utilized for the data logical qubit, that is, an additional full code block of ancilla qubits. The ancilla state preparation, is followed by the application of CNOT gates for parity-check readout. Finally, the error syndrome information is extracted from the parity-check values stored on the ancilla qubits. For the [[7,1,3]] code, the protocol further leverages the transversallity of the logical CNOT gate for CSS codes with $k=1$ logical qubit, meaning that all CNOT gates can be applied in parallel between the corresponding physical qubits of the two logical data and ancilla qubits.}

To prevent errors in faulty gadgets from propagating into harmful multi-qubit errors, Steane proposed expanding the ancilla register such that each ancilla qubit interacts transversely with only one data qubit~\cite{Steane1996_FTQEC,Steane1997_ActiveStab}, a design similar to the fault-tolerant scheme of DiVincenzo and Shor~\cite{DiVicenzoShor1996}. Unlike the latter which relies on cat-state preparation for the ancilla, Steane error correction (Steane EC) employs a fault-tolerant preparation of logical ancilla states. Leveraging the CSS structure of the [[7,1,3]] code, Steane EC simplifies stabilizer measurements into two steps: syndromes of the same type (either $X$ or $Z$) are extracted simultaneously. Each step involves seven parallel CNOT gates, applied transversally between the data and ancilla qubits, reducing the circuit depth of each step to one.

As illustrated in Fig.~\ref{fig:syndrome_extraction_circuits}a, Steane EC requires the preparation of encoded ancilla states, namely, $\ket{0}_{\rm L}$ for $X$-check and $\ket{+}_{\rm L}=H^{\otimes 7}\ket{0}_{\rm L}$ for $Z$-check, where $H$ is the single-qubit Hadamard gate. The logical state preparation must be done fault-tolerantly, which we will discuss in more details in the next section. Errors are then transferred from the data qubits via transversal CNOT gates between the data block and the ancilla block, followed by measurements of the ancilla qubits.

We now illustrate the conversion the measurement outcomes to corresponding syndrome for the $Z$-check; the process for the $X$-check is analogous.  In the absence of errors, after applying transversal CNOT gates, the ancilla state is $\ket{+}_{\rm L} =(\ket{0}_{\rm L}+\ket{1}_{\rm L})/\sqrt{2}$ . Measurements on all ancilla qubits then collapse this state to one of the 16 codewords defined in Eqs.~\ref{eq:logical0} and~\ref{eq:logical1}, producing a measurement bit string $b$. An $X$ error on a data qubit flips the corresponding bit in $b$, and the $Z$-type parity-check matrix $H_Z$ is used to determine the associated $Z$ syndrome as follows:
\begin{equation}
   s^Z =  H_{Z}\cdot b^T \mod 2,
\end{equation}
For instance, consider a bit-flip error on the second data qubit, resulting in the $Z$-check measurement bit string $b=0100000$. With $H_Z$, the corresponding syndrome is calculated as $s^Z=110$. Referring back to Table~\ref{tab:s2LUT}, this syndrome indicates that the required recovery operation is indeed $X_2$. This demonstrates that the critical information for determining the error syndrome lies in the parity information encoded in the measurement bit string, rather than the specific bit string itself.

\begin{figure*}[t!]
  \centering 
  \includegraphics[width=1\textwidth]{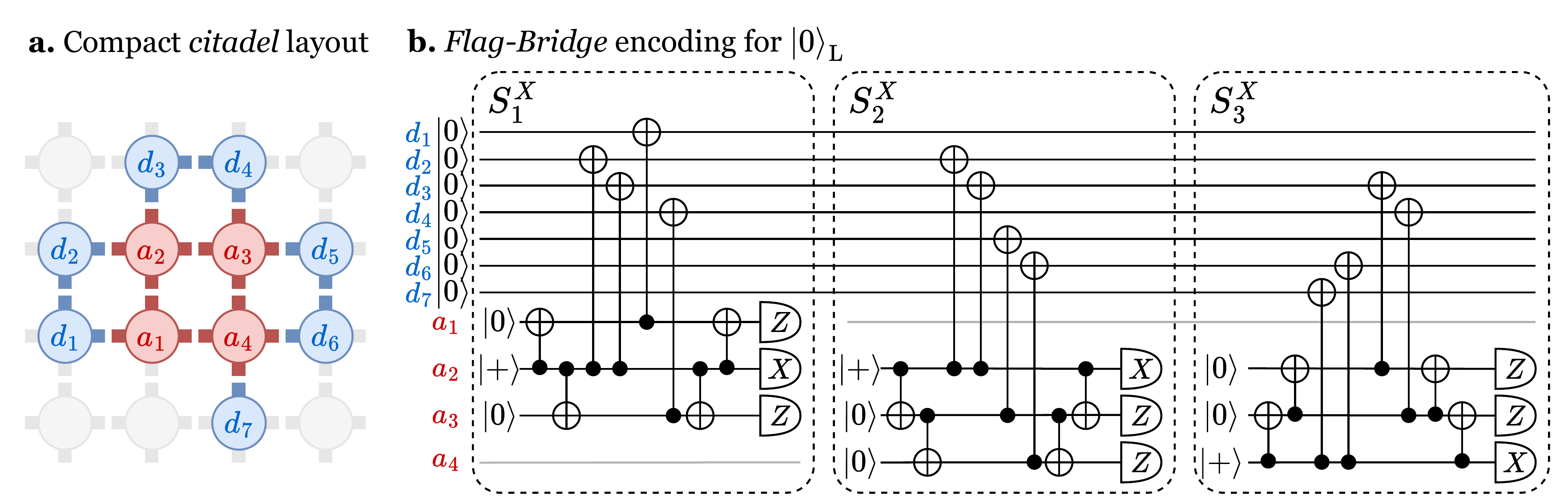}
  \caption{\textbf{Compact parity-check encoding with \textit{Flag-Bridge} qubits on a 2D grid.} a) \textit{Citadel}-like layout places data qubits (blue) around ancilla qubits (red), fitting neatly onto a $4\times 4$ lattice. This design uses a total of four ancilla qubits, which are reused for each stabilizer measurement. b)  \textit{Flag-Bridge} encoding circuit prepares the $\ket{0}_{\rm L}$ state of the [[7,1,3]] code by measuring the stabilizers $S_1^X,S_2^X,$ and $S_3^X$ sequentially. It employs Circuit 3 from Fig.~\ref{fig:syndrome_extraction_circuits}b, with one syndrome qubit and two flag qubits assigned per stabilizer.}
  \label{fig:FBEncLayout_Circuit}
\end{figure*}

\subsection{Flag-Bridge error-correction}\label{subsec:flagEC}

In contrast to Steane EC, a resource-efficient alternative method was proposed to perform FT error-correction \cite{Yoder2017surfacecodetwist, ChaoReichardt2018_twoextraqubits} that allows a significant reduction in the number of ancilla qubits required for syndrome extraction. This approach adds one or a few extra ancilla qubits, known as \emph{flag qubits}, to detect and prevent the propagation of high-weight errors that can arise during syndrome extraction.

For example, let us consider Circuit 1 in Fig.~\ref{fig:syndrome_extraction_circuits}b, which implements a weight-4 $X$-type parity-check. In this circuit, $Z$ errors on the data qubits propagate through the four CNOT gates to the syndrome qubit, flipping its initial $\ket{+}$ state if an odd number of $Z$ errors are present. Consequently, measuring the syndrome qubit in the $X$-basis reveals the parity of the $XXXX$ operator on the data qubits. At the same time, an $X$ error on the syndrome qubit occurring after the middle two CNOT gates can propagate back to the data qubits, potentially causing a harmful weight-2 error. This can be addressed with an extra flag qubit with two syndrome-to-flag CNOT gates added around the middle two CNOT gates. The modification ensures that any such $X$ error propagates to the flag qubit, flipping its initial $\ket{0}$ state, which is subsequently detected by the flag measurement in the $Z$-basis. With a single flag qubit, the modified syndrome extraction circuit limits errors on the outgoing data block to at most one qubit, satisfying the FT condition.

% \arb{ The primary role of flag qubits is to signal the presence of high-weight errors emerging from lower-weigh faults. In the context of the [[7,1,3]] code, those harmful errors arise from single faults occurring on the syndrome qubit that can propagate into more than one error within the encoded data block and remain undetectable by syndrome qubit measurements. These errors violate the level-1 fault-tolerance conditions of outputting at most a state with one single error. Thanks to flag qubits, the presence of harmful errors is detected before they escalate.} \arb{For example, in Fig.~\ref{fig:syndrome_extraction_circuits}b, all Circuit 1-3 represent weight-4 $X$-type parity-check readout circuits following a level-1 FT flag-based design.} \arb{Syndrome extraction is performed sequentially, by implementing one weight-4 parity-check at a time. The CNOT gates propagate errors from the data qubits to the syndrome qubit either by direct coupling or by entangling first with a flag qubit and this one with the syndrome qubit using additional CNOT gates. The CNOT gates between the syndrome and flag qubits allow dangerous errors to trigger flag qubits that are detected upon measurement, signaling the presence of high-weight errors on that circuit implementation.}

In addition to preserving fault tolerance, flag qubits can serve as bridges to mediate interactions between data and syndrome qubits. This role is particularly useful for mapping syndrome extraction circuits onto hardware with limited connectivity~\cite{LaoAlmudever2020}. For instance, in Fig.~\ref{fig:syndrome_extraction_circuits}b, Circuit 1 cannot be mapped onto a 2D grid with only nearest-neighbor connectivity because the syndrome qubit must interact with five other qubits.  However, by incorporating \textit{Flag-Bridge} qubits and optimizing the placement of entangling gates, Circuits 2 and 3 ensure that each qubit interacts with at most three others, making them suitable for this 2D grid topology.

The use of \textit{Flag-Bridge} qubits offers significant advantages by enabling fault-tolerance with minimal overhead, while ensuring compatibility with constrained topologies. Moreover, as we will demonstrate later, flag measurement outcomes can be leveraged to adjust the balance between performance and efficiency. This approach, referred to as \textit{\textit{Flag-Bridge} error-correction} (FB EC), and the associated syndrome extraction circuits, will serve as a central focus of our study.

% \arb{To be done: add discussion on how stabilizer measurement is done in practice, using Circuit 3 with 1 syndrome and 2 flags, refer to page 6 in October version.}

% \arb{Comment: I think we need to emphasize more the advantage of the 1 flag and the 2- flag, beyond the compact arrangement and less qubits. Because that main difference is what makes the 2-flag "powerful" for our purposes.See circuti Fig 4b in https://arxiv.org/pdf/2409.04628}

%\begin{figure}[h!]
%  \centering[width=0.8\columnwidth]{Quantiniuum_encodincircuit.png}
%  \caption{Encoding circuit for logical $|0\rangle$- Goto-picture taken from Quantiniuum- a new one willl be made}
%  \label{fig:quantiniuum}
%\end{figure}

\section{FT encoding on a 2D grid}\label{Sec:enc}
With the options for fault-tolerant error correction outlined, we now focus on the critical first step: preparing a logical state in a fault-tolerant manner in a practical setting. To address connectivity constraints in various near-term devices, we describe in this section two fault-tolerant encoding methods suitable for a 2D grid topology. We also detail the additional resources required to ensure compatibility with the FTEC protocols presented above.

\begin{figure*}[t!]
  \centering \includegraphics[width=1\textwidth]{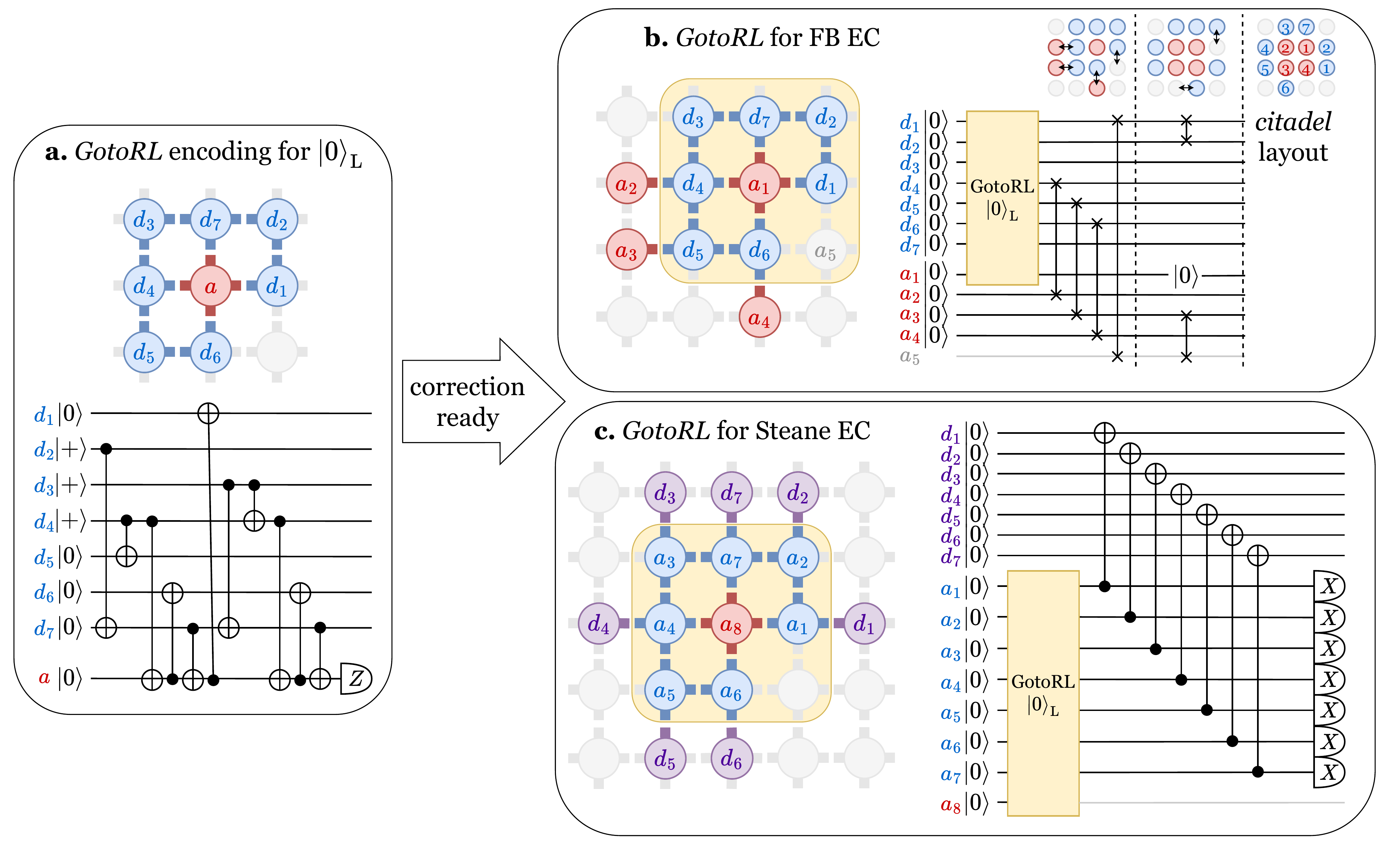}
  \caption{\textbf{Compact verification-based encoding on a 2D grid.} a) Compact layout and circuit for \textit{GotoRL} encoding were inspired by Goto's efficient-verification method~\cite{Goto2016} and discovered using a Reinforcement Learning agent~\cite{zen2024quantumcircuitdiscoveryfaulttolerant}. This approach uses a single ancilla qubit (red) for verification and as a bridge between the data qubits (blue). b) Preparing \textit{GotoRL} encoding for FB EC involves applying six SWAP gates to rearrange the qubits into the \textit{citadel} layout. The exact qubit order need not match perfectly, as discussed in the main text. c) To prepare \textit{GotoRL} encoding for Steane EC, we use the $X$-check circuit in Fig.~\ref{fig:syndrome_extraction_circuits}a. Specifically, the process repurposes the prepared $\ket{0}_{\rm L}$ state as a logical ancilla and employs transversal CNOT gates to transfer the logical state to a new set of data qubits (purple).}
\label{fig:GotoRLEncLayout_Circuit}
\end{figure*}

\subsection{Parity-check encoding with flag qubits}

% \arb{The most general method for preparing a logical state of stabilizer codes involves projecting the state onto the common $+1$ eigenspace of the stabilizers. This is done by applying the parity-check readout circuits from Fig.~\ref{fig:syndrome_extraction_circuits}b. Upon measurement on the ancilla qubits, the initial state of the data qubits gets projected onto an eigenstate associated with a particular ancilla measurement outcome.  Thus, to prepare a logical state within the simultaneous $+1$ eigenspace of the stabilizers, we measure all the stabilizers and post-select for even-parity outcomes.}

The most general method for preparing a logical state of a stabilizer code involves directly measuring its stabilizers. These measurements project the initial state of the data qubits onto an eigenstate corresponding to the obtained outcomes. To prepare a logical state within the simultaneous $+1$ eigenspace of the stabilizers - the \textit{code space} - we selectively post-process and retain only even-parity outcomes.

In particular, we can verify that the $\ket{0}_{\rm L}$ state, which is also the $+1$ eigenstate of the logical $Z$ operator, can be prepared by applying the corresponding projectors with eigenvalue $+1$ onto an arbitrary state:

\begin{equation}
\ket{0}_{\rm L} = \frac{1}{\mathcal{N}} \left( 1 + Z_{\rm L} \right) \prod_{S_i \in \mathcal{S}} \left( 1 + S_i \right) \ket{\psi},
\end{equation}
where $\mathcal{S}$ denotes the stabilizer group of the code and $\mathcal{N}$ is the normalization factor.

For the [[7,1,3]] code, the all-zero initial state $\ket{0}^{\otimes 7}$ already satisfies the even-parity condition for the three $Z$-type stabilizers and the logical $Z$ operator. Therefore, preparing the logical $\ket{0}_{\rm L}$ state only requires measuring the $X$-type stabilizers, $S_1^X$, $S_2^X$, and $S_3^X$, and selecting the $+1$ eigenvalue outcomes as follows:
\begin{equation}
\ket{0}_{\rm L} = \frac{1}{\sqrt{8}} \left( 1 + S_3^X \right) \left( 1 + S_2^X \right) \left( 1 + S_1^X \right) \ket{0}^{\otimes 7}.
\end{equation}
In practice, stabilizers are measured using syndrome qubits, with fault tolerance ensured by flag qubits, all integrated into the FT parity-check circuits shown in Fig.~\ref{fig:syndrome_extraction_circuits}b.

Previously proposed mappings onto a 2D grid topology using \textit{Flag-Bridge} qubits require a $5\times5$ lattice and six ancilla qubits~\cite{LaoAlmudever2020}. In contrast, we introduce a more space- and resource-efficient \textit{citadel}-like layout. As shown in Fig.~\ref{fig:FBEncLayout_Circuit}a, our design fits neatly within a $4\times4$ lattice and uses only four ancilla qubits for the entire code by using Circuit 3 from Fig.~\ref{fig:syndrome_extraction_circuits}b. Fig.~\ref{fig:FBEncLayout_Circuit}b further depicts the sequential execution of three $X$-type parity-check circuits, with three out of the four ancilla qubits activated per stabilizer. Since the $Z$-type parity-check circuits share the same connectivity requirements, the \textit{citadel} layout is directly compatible with the full FB EC protocol, without the need for any additional qubit reconfiguration. In essence, this \textit{Flag-Bridge} encoding is inherently correction-ready.

% Upon completing the encoding circuit, the resulting syndrome and flag strings have lengths of three and six, respectively. 

An important aspect of this encoding method is its probabilistic nature. Starting from the all-zero initial state $\ket{0}^{\otimes 7}=[(\ket{+}+\ket{-})/\sqrt{2}]^{\otimes 7}$, each $X$-type stabilizer measurement has an equal probability of yielding an odd or even parity. As the result, even in the absence of faulty gadgets, the probability of preparing the state in the codespace - achieving even-parity outcomes for all three $X$-type stabilizers - is just $(1/2)^3=1/8$. At first glance, this might suggest the need for a post-selection process to ensure proper state preparation, potentially leading to significant increase in time overhead. Fortunately, we observe that the probabilistic projections introduce only $Z$ errors. As noted in Ref.~\cite{Goto2016}, $Z$ errors are harmless to $\ket{0}_{\rm L}$ of the [[7,1,3]] code, thanks a special property of quantum codes called  \emph{error degeneracy}. More specifically, we can verify that, any $Z$ error, regardless of its weight, can be reduced to either the identity or a weight-1 error through some composition with the stabilizers $\{S_1^Z,S_2^Z,S_3^Z\}$ and the logical operator $Z_{\rm L}$. Since these additional single-qubit $Z$ errors are correctable during a subsequent error-correction cycle, post-selection is not required and non-trivial syndrome measurements can be accepted without impacting the shot efficiency of the encoding process (see below).

\begin{table}[t!]
\begin{center}
    \begin{tabular}{|c|c|c|c|c|c|} 
        \hline Encoding & EC& \# Anc. & \# Enc.  & \# Extra & Layout \\
        method & scheme & qubits & CNOTs &  CNOTs & \& circuit\\
        \hline \hline
        \textit{Flag-Bridge} & FB & 4 & 24 & 0 & Fig.~\ref{fig:FBEncLayout_Circuit} \\
        \textit{GotoRL} & FB & 5 & 11 & 18 & Fig.~\ref{fig:GotoRLEncLayout_Circuit}a-b \\
        \textit{GotoRL} & Steane & 8 & 11 & 7  & Fig.~\ref{fig:GotoRLEncLayout_Circuit}a-c \\
        \hline
    \end{tabular}
    \caption{Summary of correction-ready encoding circuits studied in this work, along with the corresponding number of ancilla qubits, CNOT gates used during encoding, and extra CNOT gates required for reconfiguration.  \textit{Flag-Bridge} encoding is inherently compatible with FB EC, requiring no additional CNOT gates. \textit{GotoRL} encoding can be adapted for either FB EC or Steane EC. The adaptation for Steane EC reduces the number of CNOT gates but increases the ancilla qubit overhead. Complete layouts and circuits for all protocols are shown in the final column.}
    \label{tab:encoding_protocols}
\end{center}
\end{table}

\subsection{Verification-based encoding}
    
The second encoding method involves using a small number of ancilla qubits for efficient fault-tolerance verification. As proposed by Goto in Ref.~\cite{Goto2016}, the approach identifies all potential harmful errors induced by a non-FT encoding circuit and determines the minimal set of parity-checks required to detect them. Notably, for the [[7,1,3]] code, Ref.~\cite{Goto2016} found that only one ancilla qubit is needed for the verification, resulting in a FT encoding circuit with remarkably low overhead. This encoding circuit has been successfully implemented in a trapped-ion architecture with all-to-all connectivity for real-time QEC demonstrations~\cite{RyanAnderson_Honeywell_RealTimeFTQEC_2021, ryananderson2022implementingfaulttolerantentanglinggates, Postler2022, Postler2024}. 

For the 2D grid topology considered in this work, we adopt a similarly resource-efficient circuit, depicted in Fig~\ref{fig:GotoRLEncLayout_Circuit}a. Inspired by Goto's efficient verification concept, the circuit was initially discovered by a reinforcement learning agent trained to design fault-tolerant logical state preparation circuits tailored to a 2D grid topology with only nearest neighbor connectivity~\cite{zen2024quantumcircuitdiscoveryfaulttolerant}. Remarkably, the circuit retains its low-overhead and uses s single ancilla qubit to also both detect harmful error and mediate interactions between data qubits. We will refer to this verification-based encoding approach as \textit{GotoRL} encoding.

Unlike \textit{Flag-Bridge} encoding, where FB EC is a natural choice as it requires no additional reconfiguration, \textit{GotoRL} encoding can be adapted to be compatible with either of the EC protocols. Fig.~\ref{fig:GotoRLEncLayout_Circuit}b illustrates the use of SWAP gates to rearrange the qubits on the 2D grid, making them compatible with FB EC. Notably, it is unnecessary to replicate the exact layout depicted in Fig.~\ref{fig:FBEncLayout_Circuit}. Instead, it suffices to ensure that each group of four data qubits associated with a stabilizer is directly connected to three ancilla qubits, enabling the use of Circuit 3 from Fig.~\ref{fig:syndrome_extraction_circuits}b. This rearrangement requires just six SWAP gates, which can be executed in two stages: the first four gates in parallel, followed by the last two gates in parallel. Additionally, the ancilla qubit employed for verification during the encoding stage must be reset. While each SWAP gate comprises three CNOT gates and introduces some performance degradation, this approach allows for a fair comparison with \textit{Flag-Bridge} encoding circuit in terms of readiness for FB EC.

For compatibility with the Steane EC protocol, Fig.~\ref{fig:GotoRLEncLayout_Circuit}c demonstrates an alternative strategy involving CNOT gates and measurements. Here, the logical state produced during the encoding process is repurposed as a logical ancilla to initialize a new logical state on a separate set of data qubits. First, all original qubits $\{d_1, \dots, d_7, a_1\}$ on the $3\times3$ lattice in \textit{GotoRL} encoding are redefined as ancilla qubits $\{a_1, \dots, a_7, a_8\}$. Then, we select new data qubits (in purple) adjacent to these redefined ancilla qubits, enabling direct pairwise application of transversal CNOT gates without requiring qubit movement, as depicted in Fig.~\ref{fig:GotoRLEncLayout_Circuit}c. This process effectively performs an $X$-check of the Steane EC protocol on the all-zero initial state $\ket{0}^{\otimes 7}$ of the new data qubits, thereby preparing $\ket{0}_{\rm L}$ up to some ultimately harmless $Z$ errors induced by probabilistic projection. Making \textit{GotoRL} encoding compatible with Steane EC requires only seven additional CNOT gates, significantly fewer than the eighteen required for FB EC (three per SWAP gate). As a result, we expect this protocol to offer superior performance, though at the cost of higher qubit overhead.

We summarize the correction-ready encoding circuits in Table~\ref{tab:encoding_protocols} detailing their respective qubit requirements and the number of entangling gate operations. In Sec.~\ref{subsec:encoding_failure}, we numerically benchmark the performance of these protocols during the encoding process. Following this, we analyze the performance of one cycle of error correction for FB EC, considering both the \textit{Flag-Bridge} and \textit{GotoRL} encoding schemes, in Sec.~\ref{subsubsec:FBEC_analysis}. Finally, we evaluate the performance of Steane EC, specifically with \textit{GotoRL} encoding, in Sec.~\ref{subsubsec:SteaneEC_analysis}.

\section{Results}\label{sec:numerical}

We evaluate the performance of the correction-ready encoding circuits through comprehensive multi-shot circuit-level noise state-vector simulations using Qulacs~\cite{Suzuki_2021_qulacs}. We use the following depolarizing model parameterized by a single physical error rate $p_{\rm phys}=p$:
\begin{enumerate}
    \item Each single-qubit gate is followed by a Pauli error $\{X,Y, \text{or } Z\}$, with probability $p$, where each error occurs with equal probability $p/3$.
    \item  Each two-qubit gate is followed by an error drawn uniformly and independently from $\{I,X,Y,Z\}^{\otimes 2} \setminus \{I\otimes I\}$ with probability $p$, giving each error a probability of  $p/15$.
    \item Each $\ket{0}$ state initialization is replaced by either $\ket{1}=X\ket{0}$ or  $i\ket{1}=Y\ket{0}$, each with probability $p/3$, giving the total probability of $2p/3$ for faulty initialization.
    \item Each $Z$-basis measurement outcome is flipped with probability $2p/3$.
    \item Preparation and measurement in the $X$-basis are achieved through application of Hadamard gates.
    \end{enumerate}

After running the simulations and collecting all measurements, we consider various ways to utilize post-selection (PS) based on syndrome and flag information to achieve fault tolerance and improve performance, at the cost of fewer accepted shots. We quantify a circuit's shot efficiency via its \emph{acceptance rate}, defined as the fraction of shots where the flag outcomes fall within a predefined set. 

% With the exception of Sec.~\ref{subsec:encoding_failure} where the specific flag sets are explicitly stated, PS refers to accepting only shots with trivial relevant measurement outcomes.

\subsection{Benchmarking isolated encoding}\label{subsec:encoding_failure}

\begin{figure}[t!]
  \centering \includegraphics[width=0.49\textwidth]{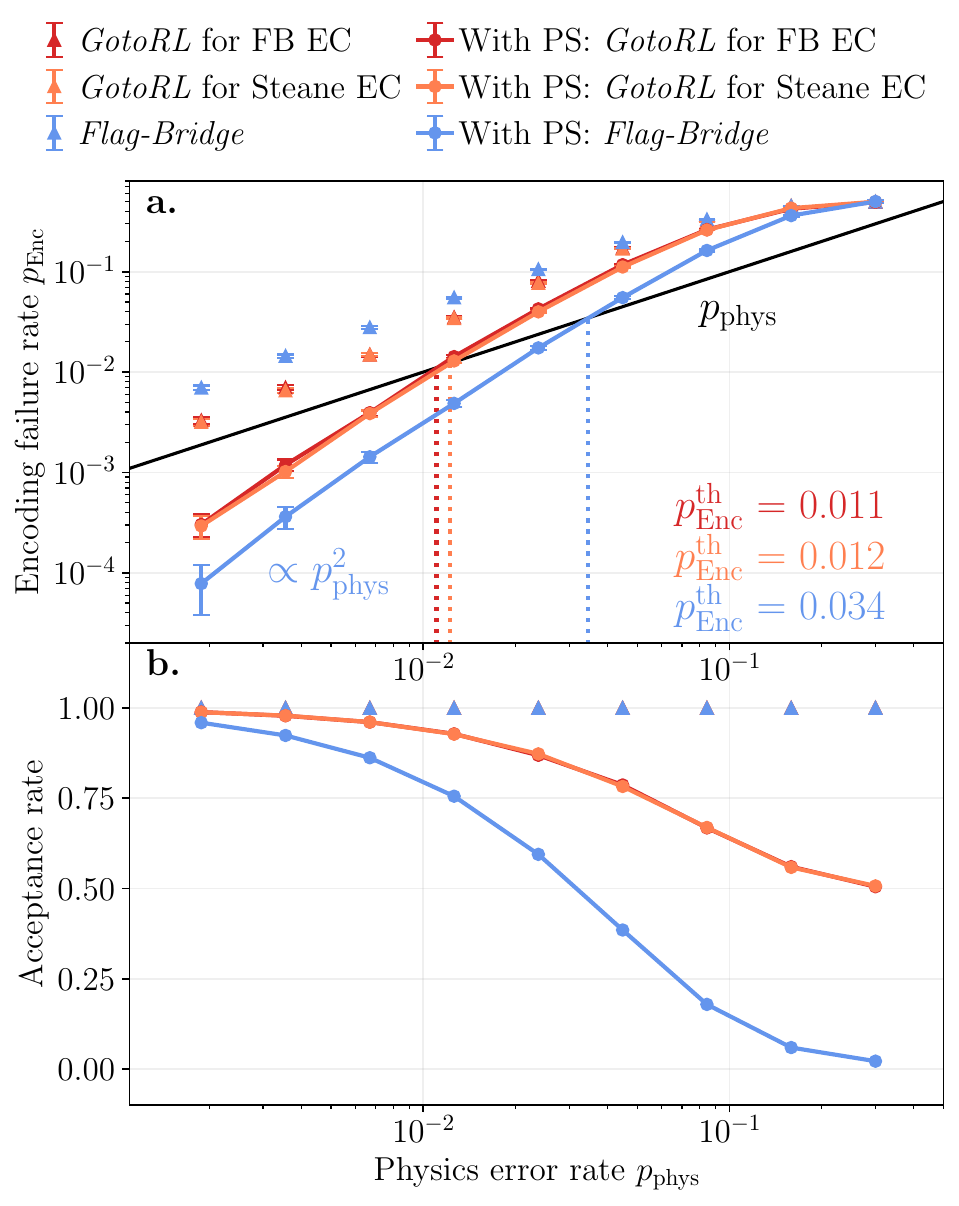}
  \caption{\textbf{Encoding-only performance and efficiency.} a) \textit{Encoding failure rate} $p_{\rm Enc}$ is plotted as a function of the physical error rate $p_{\rm phys}$ for the protocols summarized in Table~\ref{tab:encoding_protocols}, with and without post-selecting for trivial flag outcomes. Post-selection with flags achieves fault-tolerant scaling, characterized by $p_{\rm Enc} \propto p_{\rm phys}^2$. The encoding pseudo-threshold, marked by the intersection with $p_{\rm Enc} = p_{\rm phys}$, is doubled for \textit{Flag-Bridge} encoding compared to \textit{GotoRL} encoding. b) The corresponding acceptance rates indicate that \textit{GotoRL} encoding achieves greater shot efficiency, maintaining acceptance rates above 50\%. Results are obtained from circuit-level noise simulations with 200,000 shots per data point.}
\label{fig:encodings_performance}
\end{figure}

\begin{figure}[t!]
\vspace{5mm}
  \centering \includegraphics[width=0.49\textwidth]{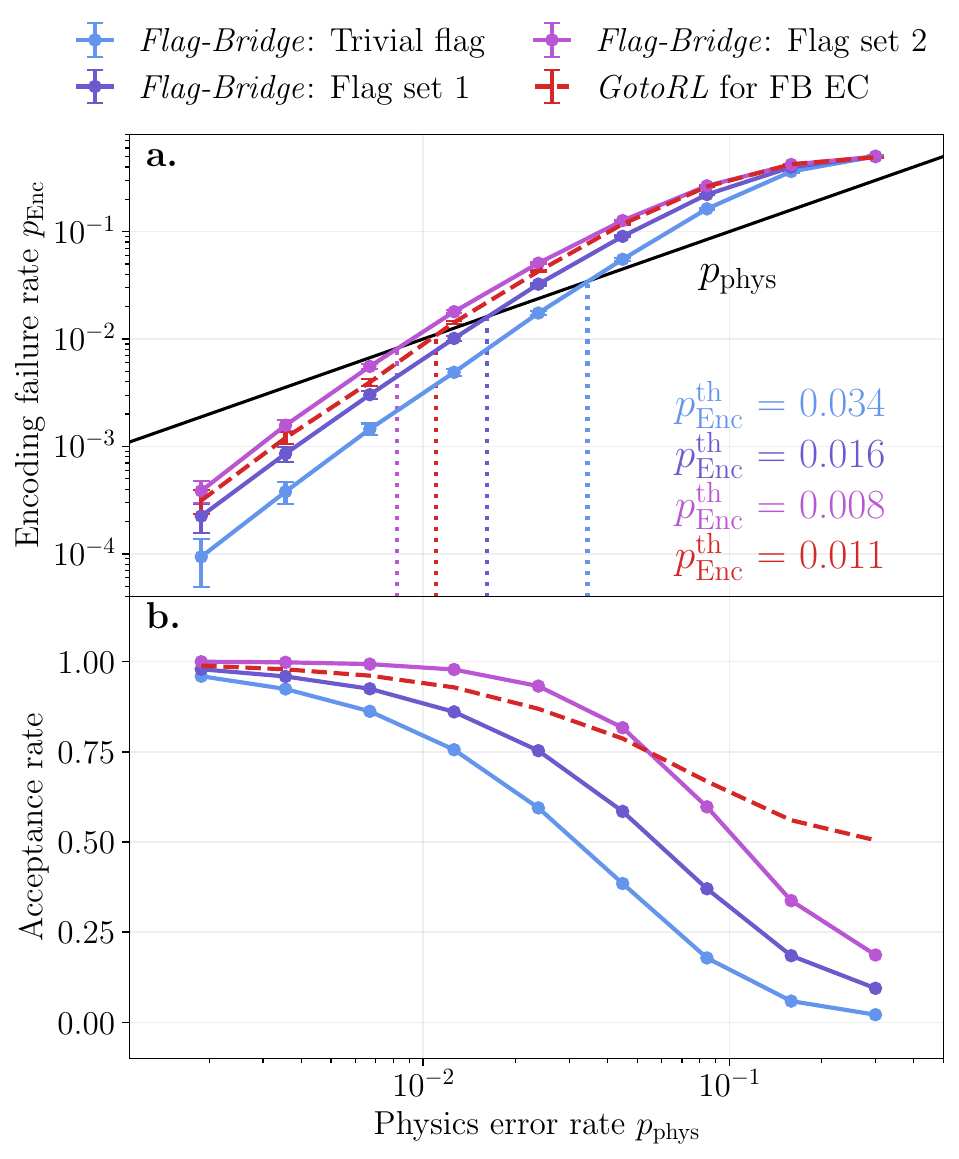}
  \caption{\textbf{Encoding-only performance and efficiency trade-off for \textit{Flag-Bridge} approach.}  a) \textit{Encoding failure rates} for different flag sets, as detailed in Table~\ref{tab:flags_breakdown}, worsen progressively as more flag patterns are accepted. Despite this, the scaling remains quadratic, demonstrating that fault-tolerant behavior can be preserved with strategic broadening of the accepted flag set (see App.~\ref{app:nontrivial_flag_patterns}). b) Corresponding acceptance rates improve significantly with broader flag sets, approaching the levels achieved by \textit{GotoRL} encoding. This highlights the flexibility of the \textit{Flag-Bridge} approach, enabling users to balance performance and efficiency according to specific requirements. Results are obtained from circuit-level noise simulations with 200,000 shots per data point.}
\label{fig:flagbridge_tradeoff}
\end{figure}

As an initial benchmark, we run the noisy encoding circuits and assess whether the encoded states contain harmful, uncorrectable errors. The [[7,1,3]] code is designed to independently correct for single-qubit $X$ and $Z$ errors. Thus, a failed encoding results an error where either the $X$-part or $Z$-part spans two data qubits. However, this failure condition simplifies further, as only two-qubit $X$ errors are harmful for the $\ket{0}_{\rm L}$ state of the [[7,1,3]] code~\cite{Goto2016}. Using this criterion, we define the \emph{encoding failure rate} as:
\begin{equation}
    p_{\rm Enc} = \lim_{N_{\rm accepted}\rightarrow\infty} \frac{N_{\rm 2Q-X\text{error}}}{N_{\rm accepted}},
\end{equation}
where $N_{\rm accepted}$ is the number of accepted shots, chosen to be sufficiently large to ensure an accurate estimate of  $p_{\rm Enc}$. We also provide error bars based on standard error of the binomial distribution
\begin{equation}
    \sigma(p_{\rm Enc}) = \sqrt{\frac{p_{\rm Enc}(1-p_{\rm Enc})}{N_{\rm accepted}}}.
\end{equation}
Since the FT protocol is specifically designed to detect harmful weight-2 errors arising from a single fault, leaving only those caused by two or more faults, $p_{\rm Enc}$ is expected to scale as $p^2$.

Fig.~\ref{fig:encodings_performance} compares the performance of the outlined correction-ready encoding circuits: \textit{Flag-Bridge}, \textit{GotoRL} for FB EC, and \textit{GotoRL} for Steane EC. For all circuits, we observe the expected transition of the scaling of $p_{\rm Enc}$ from $p$ to $p^2$, when moving from a non-FT protocol (without PS) to a FT protocol (with PS). To evaluate the performance more concretely, we define the \emph{encoding pseudo-threshold} as the value of $p$ at which
\begin{equation}
    p_{\rm Enc}(p) = p,
\end{equation}
indicating the point below which an encoded qubit outperforms a physical qubit. Using this metric, we find that \textit{GotoRL} encoding prepared for Steane EC performs marginally better than that for FB EC. This is expected, as reconfiguration for Steane EC requires fewer entangling gates (see Table~\ref{tab:encoding_protocols}), leading to fewer opportunities for errors to accumulate.

Notably, the \textit{Flag-Bridge} circuit significantly outperforms both \textit{GotoRL} circuits, achieving an encoding pseudo-threshold more than double that of the others (increasing from 1.2\% to 3.4\%). The improvement stems from the use of flag qubits after each parity-check, enabling early error detection. In contrast, \textit{GotoRL} encoding executes more CNOT gates before measuring the verification qubit, increasing the chance for errors to accumulate into false positives. However, the lower failure rate comes at the expense of the acceptance rate: for large $p$, the acceptance rate of \textit{Flag-Bridge} encoding can approach 0\%, whereas \textit{GotoRL} encoding maintains an acceptance rate above 50\%. For near-term devices where noise levels remain high, the latter choice appears to be better suited when shot efficiency is a critical factor.

Unlike \textit{GotoRL} encoding, which uses only a single flag qubit, the \textit{Flag-Bridge} circuit employs six flag qubits, resulting in $2^6$ possible flag patterns. This design enables a tunable trade-off between performance and acceptance rate by expanding the set of accepted flag patterns in post-selection. Thus far, we have reported the encoding failure rate $p_{\rm Enc}$ for the trivial flag pattern `00 00 00'. Table~\ref{tab:flags_breakdown} categorizes additional flag patterns based on their individual contributions to \(p_{\rm Enc}\). Specifically, `Flag set 1' includes the trivial pattern plus four others, while `Flag set 2' extends `Flag set 1' by adding five more, each with progressively higher individual $p_{\rm Enc}$. These selected patterns plausibly result from at most one fault, while those clearly indicating multiple faults are excluded. For instance, a pattern like `00 10 01' suggests at least two faults—one in the $S^X_2$ plaquette and another in the $S_3^X$ plaquette, and is therefore omitted.  

Figure~\ref{fig:flagbridge_tradeoff} illustrates the achievable range of shot efficiency when considering the flag sets defined in Table~\ref{tab:flags_breakdown}, along with the corresponding degradation in $p_{\rm Enc}$. Notably, in some cases, shot efficiency can even surpass that of \textit{GotoRL} encoding. This underscores the versatility of the \textit{Flag-Bridge} circuit, which allows for dynamic tuning of the performance-efficiency trade-off during post-processing by adjusting the accepted flag set without requiring any physical modifications to the circuit itself.

\begin{table}[t!]
\begin{center}
    \begin{tabular}{|c|c|c|c|c|c|}
        \hline 
        \multicolumn{3}{|c}{} & \multicolumn{3}{|c|}{$p_{\rm phys}$} \\
        \hline
        \multicolumn{2}{|c|}{} & Flag pattern & 1.3$\times 10^{-2}$ & 2.4$\times 10^{-2}$ & 4.5$\times 10^{-2}$ \\
        \hline \hline
        \multirow{10}{*}{\rotatebox{90}{Flag set 2 (+ correction)}}  
            & \multirow{5}{*}{\rotatebox{90}{Flag set 1}} 
            & 00 00 00  & 0.005 & 0.017 & 0.055 \\ \cline{3-6}
            && 10 00 00  & 0.046 & 0.089 & 0.160 \\
            && 01 00 00  & 0.051 & 0.090 & 0.160 \\
            && 00 01 00  & 0.050 & 0.088 & 0.157 \\
            && 00 00 10  & 0.045 & 0.087 & 0.154 \\ \cline{2-6}
            && 00 11 00  & 0.075 & 0.138 & 0.205 \\
            && 00 00 11  & 0.066 & 0.118 & 0.203 \\
            && 11 00 00*  & 0.082 & 0.141 & 0.232 \\
            && 00 10 00*  & 0.077 & 0.123 & 0.215 \\
            && 00 00 01*  & 0.074 & 0.128 & 0.224 \\
        \hline
        \multicolumn{2}{|c|}{} & 00 10 01*  & 0.229 & 0.266 & 0.332 \\
        \multicolumn{2}{|c|}{\rotatebox{90}{\parbox{0cm}{Multi \\ faults}}} 
            & \vdots  & \vdots & \vdots & \vdots \\
        \hline
    \end{tabular}
    \caption{Flag pattern breakdown supporting the performance and efficiency trade-off analysis in Fig.~\ref{fig:flagbridge_tradeoff}. The table presents the \textit{encoding failure rate} $p_{\rm Enc}$ per flag pattern for three physical error probabilities $p_{\rm phys}$, \textit{Flag-Bridge} encoding. Each $X$-type stabilizer is associated with one syndrome and two flags, resulting in six flags in total for three stabilizers. The trivial flag pattern `00 00 00,' corresponding to the blue curve in Fig.~\ref{fig:flagbridge_tradeoff}, achieves the best performance but the lowest shot efficiency. Flag sets 1 and 2 incorporate additional flag patterns with comparable $p_{\rm Enc}$ per pattern. Fault tolerance is ensured by limiting the analysis to patterns that imply only a single fault per protocol; patterns suggesting multiple faults (non-trivial outcomes in more than one stabilizer) are excluded. A few patterns, marked with an asterisk (*), require specific corrections to maintain fault-tolerance, as detailed in App.~\ref{app:nontrivial_flag_patterns}.}     \label{tab:flags_breakdown}
\end{center}
\end{table}

\subsection{Benchmarking encoding + error-correction}\label{subsec:FTEnc_1ECcycle}

Thus far, our benchmarking of the FT encoding circuits has assumed noiseless error-correction. Next, we extend our analysis to include a noisy FT EC cycle and evaluate performance using the \emph{logical error rate} $p_{\rm L}$ and the \emph{pseudo threshold} defined as the physical error rate $p$ such that $p_{\rm L}(p)=p$. We simulate both encoding and EC circuits and record the resulting error $E$ on the data qubits. Based on the flag and syndrome outcomes, a look-up-table (LUT) decoder is employed to determine the recovery operation $R$. 

A logical error occurs when $[R \cdot E, Z_{\rm L}] \neq 0$, where $Z_{\rm L} = Z_1 Z_4 Z_7$ for the [[7,1,3]] code. This criterion becomes unreliable when syndrome measurements are noisy, as it assumes that $R \cdot E$ has already returned the state to the code space. Consequently, this method overestimates the \textit{true} logical error rate by including non-$X_{\rm L}$ errors, such as $X_1$, $X_4$, and $X_7$. Unless stated otherwise, the logical error rate $p_{\rm L}$ refers to the \textit{estimated} $p_{\rm L}$, in contrast to the \textit{true} $p_{\rm L}$, which accounts solely for errors that contain the $X_{\rm L}$ operator. 
\begin{eqnarray}
    \text{\textit{true }} p_{\rm L}&:&R\cdot E\sim X_{\rm L} \nonumber \\
    \text{\textit{estimated }} p_{\rm L}&:&R\cdot E\sim X_{\rm L},X_1,X_4,X_7,X_1X_2,... \label{eq:logical_error_rate}
\end{eqnarray}
where $\sim$ represents stabilizer-equivalence. This distinction will be revisited later when we examine the scaling behavior of $p_{\rm L}$ with respect to the physical error rate. 

% \arb{[I think you should include what is the additional criteria for the true then; commutator and...develop a bit more here about the criteria for the true logical error estimates. I will even numerate or bullet point the conditions so taht you can refer to them later on the text.  ]}\nam[]{It is already defined as containing the logical operator $X_{\rm L}$.}

A straightforward way to incorporate error correction is to append a full EC cycle after encoding. This approach, which we call the \textit{bare protocol}, executes the FT encoding and FT EC circuits sequentially, with each stage processing information independently. However, recognizing that encoding and error correction share overlapping circuit components, we also consider a more efficient approach: the \textit{hybrid protocol}, where the overlapping portion is executed only once, similar to the extended rectangle in Ref.~\cite{aliferis2005quantumaccuracythresholdconcatenated}. This allows information collected during encoding to be directly leveraged for error correction. In the following, we analyze the performance of the three correction-ready encoding circuits from Table~\ref{tab:encoding_protocols} with their respective EC schemes.

% \arb{Comment: I think that we should explictly differentiate the bare and hybrid in the text.
% See below}

% \arb{Another important consideration is whether the combination of different level-1 fault-tolerant circuits within the entire QEC implementation functions effectively as a level-1 gadget. We examine two possible configurations: (a) the \textit{bare} approach, where the level-1 encoding and error-correction circuits are executed sequentially, with each part processing information independently, and (b) the \textit{hybrid} protocol, where there is an overlap in the information collected from both parts}

% \arb{i) Bare FB EC \\
% - Goto  encoding 
% - \textit{Flag-Bridge} encoding }

% \arb{ii) Hybrid FB Encoding  \\
% - \textit{Flag-Bridge} encoding }

% \arb{iii) Hybrid Steane EC 
% -Goto encoding +Steane }

\subsubsection{FB EC with \textit{GotoRL} and \textit{Flag-Bridge} encodings}\label{subsubsec:FBEC_analysis}

\begin{figure}[t!]
    \vspace{-7mm}
  \centering \includegraphics[width=1.1\columnwidth]{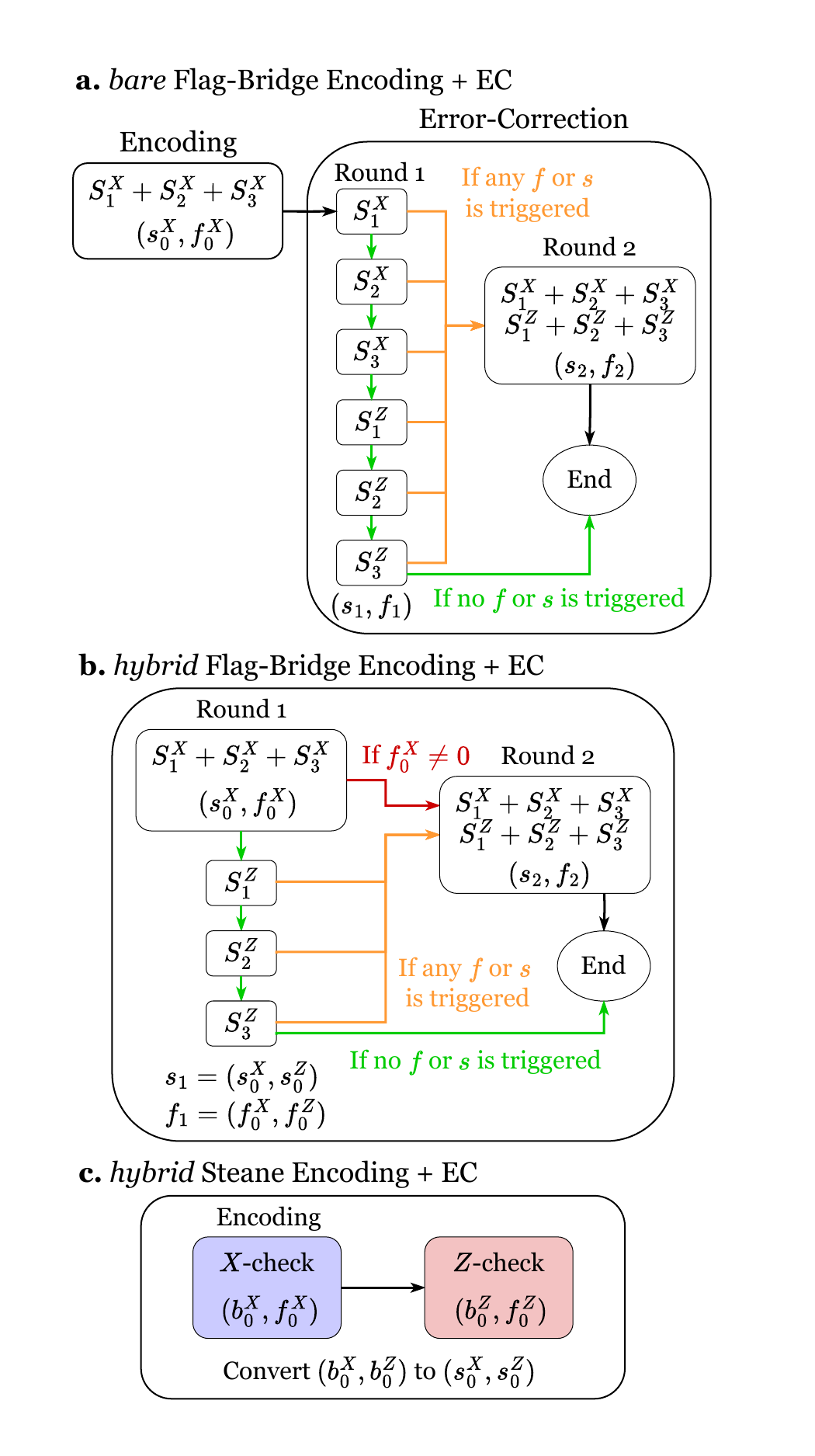}
  \caption{\textbf{Fault-tolerant encoding + EC flowcharts.} a) The \textit{bare} \textit{Flag-Bridge} protocol begins with a FT encoding, followed by a full cycle of FT EC, consisting of two rounds to re-extract syndromes in case faults are detected. Measurement outcomes $(s_1, f_1, s_2, f_2)$ from the EC stage are used to correct errors, while flag outcomes from the encoding stage, $f_0^X$, are used for additional post-selection. b) A \textit{hybrid} \textit{Flag-Bridge} protocol combines encoding and EC into a single, streamlined protocol. Syndrome information $s_0^X$ from the encoding stage is leveraged to assist with error correction. Due to the probabilistic nature of encoding, non-trivial $s_0^X$ outcomes are permitted in the first round.  c) A similar \textit{Hybrid} approach can be adapted for Steane EC using the $X$-check circuit from Fig.~\ref{fig:syndrome_extraction_circuits}a. Here, the measurement strings ($b_0^X,b_0^Z$) are converted to ($s_0^X,s_0^Z$) using the parity-check matrices in Eq.~\ref{eq:parity_check_matrix}.}
  \label{fig:EncEC_flowchart}
\end{figure}

\begin{figure}[t!]
  \centering \includegraphics[width=\columnwidth]{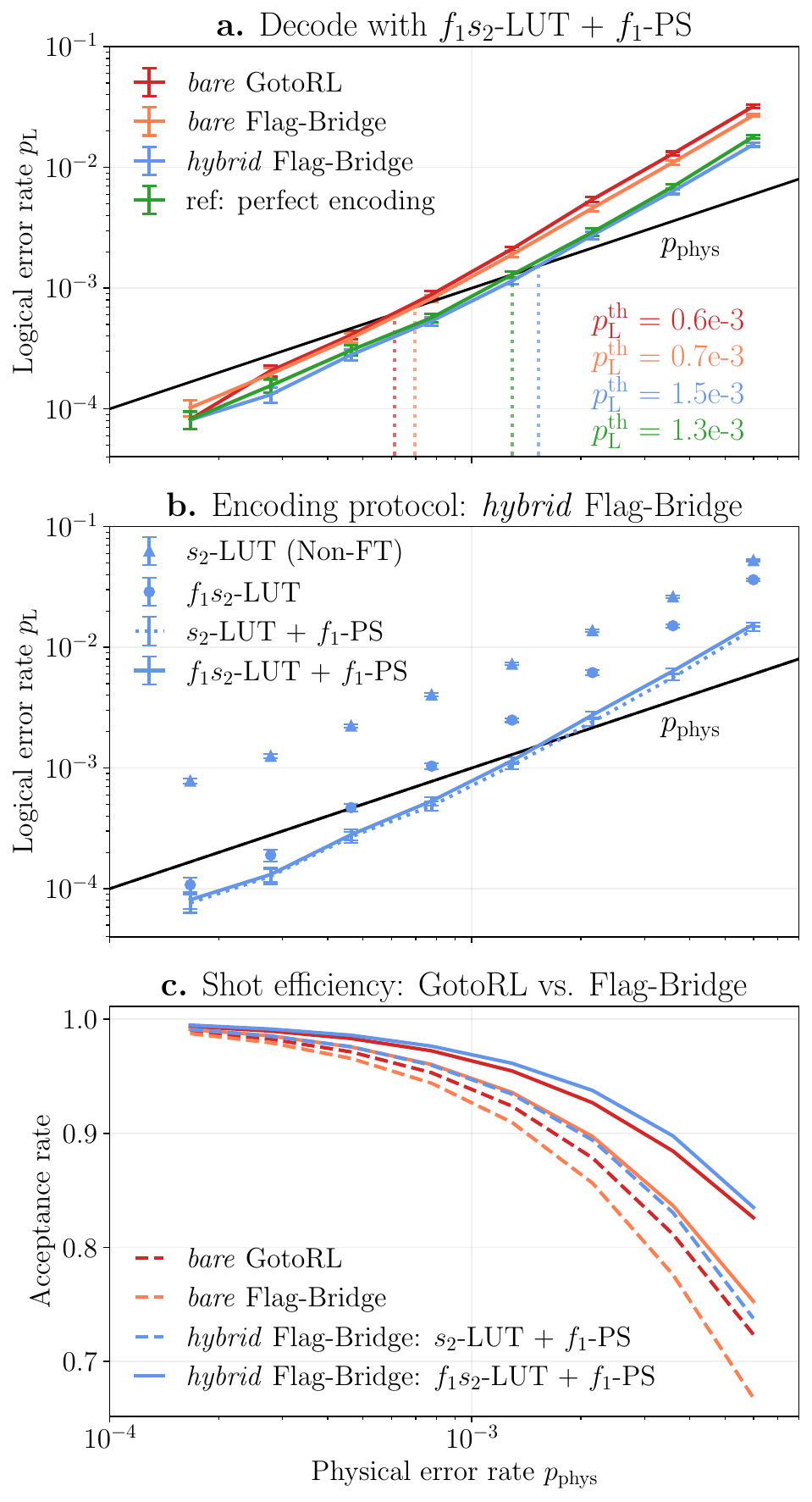}
  \caption{\textbf{Encoding + EC performance and efficiency for FB EC.} a) Estimated \textit{logical error rate} $p_{\rm L}$ is plotted against the physical error rate $p_{\rm phys}$ for protocols prepared for FB EC. With the optimal decoding strategy, the \textit{hybrid} \textit{Flag-Bridge} protocol outperforms the \textit{bare} protocol and achieves comparable performance to a noisy EC cycle assuming perfect encoding. b) For the \textit{hybrid} \textit{Flag-Bridge} protocol, optimal performance is achievable using a suboptimal decoding strategy that uses $f_1$ solely for post-selection. c) Corresponding acceptance rates demonstrate the efficiency of the protocols. Error bars indicate a 95\% confidence level and are derived from circuit-level noise simulations, with the number of shots, from left to right, being [1800, 1600, 1400, 1200, 1000, 300, 240, 220]$\times 10^3$.}
\label{fig:FBEC_analysis}
\end{figure}

\textit{Bare protocol:} We first treat encoding and error-correction separately. Syndrome and flag measurement outcomes for the encoding step are stored as $(s_0,f_0)$. For \textit{GotoRL} encoding, $s_0$ is empty and $f_0$ contains a single outcome from the verification qubit previously seen in Fig.~\ref{fig:GotoRLEncLayout_Circuit}a. For \textit{Flag-Bridge} encoding, $(s_0,f_0) = \left(s_0^X, f_0^X\right)$ representing the measurement outcomes obtained from executing three $X$-type parity-check circuits. Fault tolerance in the encoding step is ensured via post-selection based on the flag outcomes $f_0$.

The encoding is followed by a FT EC scheme with flag qubits based on Ref.~\cite{LaoAlmudever2020}. A complete EC cycle, as illustrated in Fig.~\ref{fig:EncEC_flowchart}a, consists of two rounds:
\begin{enumerate}
    \item In the first round, each stabilizer circuit $S_i\in \mathcal{S}$ is executed sequentially, with the syndrome and flag outcomes recorded as binary strings $s_1$ and $f_1$, respectively. If any flag outcome is non-trivial or the syndrome outcome differs from previously recorded syndrome information (from encoding or a prior cycle), the round is terminated early, and zeros are appended to $f_1$ to account for the skipped stabilizer circuits.
    \item If the first round completes without early termination, the second round is skipped, and we set $s_2=s_1$ indicating no detected faults. If faults were detected in the first round, all stabilizer circuits are executed without interruption, and the syndrome outcomes are stored in $s_2$.
\end{enumerate}
Note that for the \textit{bare} protocol with \textit{Flag-Bridge} encoding, partial syndrome information $s_0^X$ during encoding step serves as a reference to determine whether the syndrome outcome $s_1$ in the outlined EC cycle has changed from the encoding process. After the EC cycle, $s_2$ and $f_1$ are used for decoding.

\textit{Hybrid protocol:} Observing that \textit{Flag-Bridge} encoding inherently executes all three $X$-type stabilizers - half of the first round of the EC cycle - we propose a \emph{hybrid} protocol that leverages the information gathered during the encoding stage. As depicted in Fig.~\ref{fig:EncEC_flowchart}b, this protocol modifies the first round, to perform the three $X$-type stabilizer circuits uninterrupted (producing $s_0^X,f_0^X$), followed by the $Z$-type stabilizer circuits executed sequentially (producing in $s_0^Z,f_0^Z$). The termination condition remains unchanged except that, after running the first three stabilizers, we check only for non-trivial flag outcomes, i.e. whether $\sum f_0^X \neq 0$. By integrating the encoding step into the first round, the measurement outcomes are redefined as $s_1=(s_0^X,s_0^Z)$ and $f_1=(f_0^X,f_0^Z)$. The rest of the protocol follows the \textit{bare} protocol exactly, after which, $s_2$ and $f_1$ are used for decoding. Unlike in the \textit{bare} protocol, $f_0^X$ is not immediately used for post-selection. Instead, it is now part of $f_1$ and will be used later to enhance the decoding process.

% \nam[]{mention the similarily between our approach and Fig 2 of https://arxiv.org/pdf/quant-ph/0504218: overlapping extended rectangle. Identify overlapping portion and combine two protocols into one compact protocol to not have to repeat that twice.}

After running the EC circuits, decoding is performed on using one of two different LUTs. The first, referred to as the $s_2$-LUT, relies solely on the syndrome information from the second round, It is the standard LUT for the [[7,1,3]] code already given in Table~\ref{tab:s2LUT}. The second, referred to as the $f_1s_2$-LUT, combines the flag information from the first round and the syndrome information from the second round (Table~\ref{tab:f1Xs2ZLUT}). Incorporating flag information is advantageous because certain $f_1$ flag patterns can uniquely identify specific two-qubit $X$ errors propagated through the first round, enabling a more accurate recovery operation and reducing the likelihood of logical errors. The details of this enhanced LUT are discussed in App.~\ref{app:FBEC_f1s2LUT}.

Decoding performance can be further improved through post-selection with flag outcomes $f_1$. When using the $s_2$-LUT, only shots with trivial flags in $f_1$ are accepted. Similarly, for the $f_1s_2$-LUT, only shots with $f_1$ patterns explicitly accounted for in the LUT are accepted. Since flag qubits are critical for ensuring fault tolerance, decoding with the $s_2$-LUT without any post-selection is expected to exhibit non-FT behavior. 

% For the same reason, failing to utilize the flag information obtained during encoding in the \emph{bare} protocol also leads to non-FT behavior. Consequently, as noted in Fig.~\ref{fig:EncEC_flowchart}a, the \emph{bare} protocol must already include post-selection using $f_0^X$.

Fig.~\ref{fig:FBEC_analysis}a compares the logical error rates of different encoding+EC protocols using the optimal decoding strategy: $f_1s_2$-LUT with post-selection on $f_1$. While the \emph{bare} FB protocol demonstrates a slight advantage over the \emph{bare} \textit{GotoRL} protocol (0.07\% versus 0.06\% pseudo-threshold), both fall short in comparison to a stand-alone EC cycle, which is equivalent to a \emph{bare} protocol with perfect (noiseless) encoding (0.13\% pseudo-threshold). Notably, the \emph{hybrid} protocol enabled by FB encoding manages to bridge the gap, achieving a performance comparable to the perfect encoding scenario. Furthermore, Fig.~\ref{fig:FBEC_analysis}c shows that the \emph{hybrid} FB approach eliminates the previously observed advantage of \textit{GotoRL} encoding in terms of shot efficiency (blue versus red curves). This significant improvement is due not only to the shorter circuit length of the \emph{hybrid} protocol, which reduces the opportunities for faulty gadgets, but also to its effective use of information gathered during encoding for error correction.

Finally, we highlight in Fig.~\ref{fig:FBEC_analysis}b the possibility of achieving performance comparable to the optimal decoding strategy using a sub-optimal one - decoding with the $s_2$-LUT and accepting only shots with trivial $f_1$ (blue dotted line)- at the cost of lower acceptance rate. This is noteworthy because the control logic of the FT EC cycle (c.f. Fig.~\ref{fig:EncEC_flowchart}), with its multiple conditional branches, could pose a challenge in hardware implementation (e.g. on an FPGA). By selecting this decoding strategy in advance, the decision tree after termination can be simplified to a straightforward abort and restart, thereby making the implementation more feasible in a practical setting.

\subsubsection{Steane EC with \textit{GotoRL} encoding}\label{subsubsec:SteaneEC_analysis}

\begin{figure}[t!]
  \centering \includegraphics[width=\columnwidth]{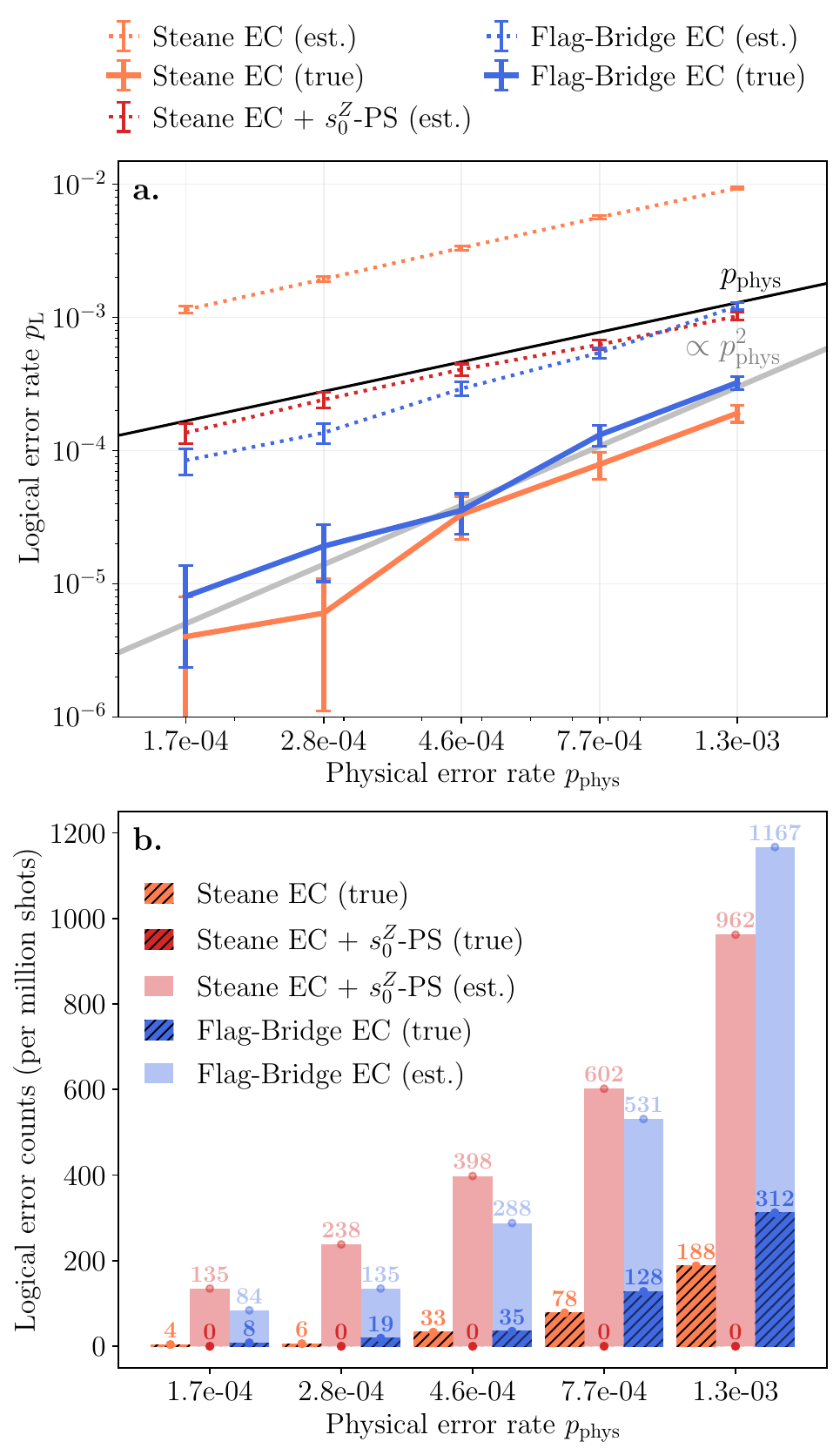}
  \caption{\textbf{Encoding + EC performance comparison between FB EC and Steane EC.} a) We plot the logical error rates $p_{\rm L}$ for \textit{hybrid} encoding-EC protocols in the low-noise regime. In terms of \textit{estimated} $p_{\rm L}$, Steane EC requires post-selection with the syndrome from the $Z$-check to match the performance of FB EC. In terms of \textit{true} $p_{\rm L}$, Steane EC without post-selection and FB EC show comparable performance, with $p_{\rm L} \propto p_{\rm phys}^2$. b) When considering the counts of \textit{true} logical errors per million shots, Steane EC with post-selection results in zero logical errors. Even without post-selection, the count of logical errors for Steane EC is approximately half that of the \textit{Flag-Bridge} approach.}
\label{fig:SteaneEC_vs_FBEC}
\end{figure}

In contrast to the previous two circuits which use FB EC, the last circuit in Table~\ref{tab:encoding_protocols} employs Steane EC, which is FT due to its use of transversal CNOT gates involving only one data qubit at a time. As seen in Fig.~\ref{fig:GotoRLEncLayout_Circuit}c, the circuit inherently incorporates the $X$-check in Steane EC. Consequently, we can implement a similarly efficient \emph{hybrid} protocol that combines encoding and EC, as depicted in Fig.~\ref{fig:EncEC_flowchart}c, for fair comparison with the \textit{hybrid} Flag-Bridge protocol.

Similarly in this protocol, the $X$-check serves a dual purpose: encoding and parity-check. During this step, we collect flag outcomes $f_0^X$ from the ancilla encoding and bit strings $b_0^X$ from the Steane parity-check. Next, we perform the standard $Z$-check to collect $f_0^Z$ and $b_0^Z$. Both $f_0^X$ and $f_0^Z$ are used for post-selection to ensure FT encoding of the logical ancilla states $\ket{0}_{\rm L}$ and $\ket{+}_{\rm L}$, respectively. The syndromes $\left(s_0^X,s_0^Z\right)$ are then computed from the bit strings $\left(b_0^X,b_0^Z\right)$ using the parity-check matrices, as outlined in Sec.~\ref{subsec:SteaneEC}. The recovery operation is determined using the standard LUT in Table~\ref{tab:s2LUT}.

\begingroup
\setlength{\tabcolsep}{5pt}
\begin{table}[t!]
\begin{center}
    \begin{tabular}{|ccccccc|} 
        \hline Index & Gate & Fault & Initial Pauli & $s$ & $f$ & Error \rule{0pt}{2.5ex} \\
        \hline \hline
		1 & CX[4,5] & -$X$ & -\,-\,-\,-$|$-$X$- & 0 & 10 & -\,-$X$$X$ \\
		1 & CX[4,5] & -$Y$ & -\,-\,-\,-$|$-$Y$- & 1 & 10 & -\,-$X$$X$ \\
		1 & CX[4,5] & $X$- & -\,-\,-\,-$|$$X$-\,- & 0 & 10 & $X$$X$-\,- \\
		1 & CX[4,5] & $X$$Z$ & -\,-\,-\,-$|$$X$$Z$- & 1 & 10 & $X$$X$-\,- \\
		1 & CX[4,5] & $Y$- & -\,-\,-\,-$|$$Y$-\,- & 1 & 10 & $X$$X$-\,- \\
		1 & CX[4,5] & $Y$$Z$ & -\,-\,-\,-$|$$Y$$Z$- & 0 & 10 & $X$$X$-\,- \\
		1 & CX[4,5] & $Z$$X$ & -\,-\,-\,-$|$$Z$$X$- & 1 & 10 & -\,-$X$$X$ \\
		1 & CX[4,5] & $Z$$Y$ & -\,-\,-\,-$|$$Z$$Y$- & 0 & 10 & -\,-$X$$X$ \\
		2 & CX[5,6] & $X$$X$ & -\,-\,-\,-$|$-$X$$X$ & 0 & 10 & -\,-$X$$X$ \\
		2 & CX[5,6] & $X$$Y$ & -\,-\,-\,-$|$-$X$$Y$ & 1 & 10 & -\,-$X$$X$ \\
		2 & CX[5,6] & $Y$$X$ & -\,-\,-\,-$|$-$Y$$X$ & 1 & 10 & -\,-$X$$X$ \\
		2 & CX[5,6] & $Y$$Y$ & -\,-\,-\,-$|$-$Y$$Y$ & 0 & 10 & -\,-$X$$X$ \\
		3 & CX[4,0] & $X$$X$ & $X$-\,-\,-$|$$X$-\,- & 0 & 10 & $X$$X$-\,- \\
		3 & CX[4,0] & $X$$Y$ & $Y$-\,-\,-$|$$X$-\,- & 0 & 10 & $Y$$X$-\,- \\
		3 & CX[4,0] & $Y$$X$ & $X$-\,-\,-$|$$Y$-\,- & 1 & 10 & $X$$X$-\,- \\
		3 & CX[4,0] & $Y$$Y$ & $Y$-\,-\,-$|$$Y$-\,- & 1 & 10 & $Y$$X$-\,- \\
        \hline
    \end{tabular}
    \caption{Summary of all single faults that lead to harmful errors on data qubits at the end of Circuit 3 in Fig.~\ref{fig:syndrome_extraction_circuits}b. For reference, the associated gate sequence is as follows: H[4], CX[4,5], CX[5,6], CX[4,0], CX[4,1], CX[5,2], CX[6,3], CX[5,6], CX[4,5], H[4]. Results are obtained from Clifford simulations detailed in App.~\ref{app:verifyFT}. For all of these faults, at least one flag outcome is non-trivial, confirming that Circuit 3 is fault-tolerant.}
    \label{tab:verifyFT_circuit3}
\end{center}
\end{table}
\endgroup

An initial comparison in the low-noise regime between these \emph{hybrid} protocols, as shown in Fig.~\ref{fig:SteaneEC_vs_FBEC}a, reveals that the \textit{estimated} logical error rate $p_{\rm L}$ for Steane EC is unexpectedly an order of magnitude higher than that of FB EC (orange dotted vs. blue dotted curves). We can enhance the performance of Steane EC by adopting a strategy similar to the \emph{hybrid} Flag-Bridge protocol, utilizing $s_0^Z$ to detect faults occurring during the procedure. Unlike the Flag-Bridge protocol's 2-round structure, which allows a second run if a fault is detected via a non-trivial syndrome $s_1$, here in Steane EC we can only use $s_0^Z$ for post-selection. Specifically, we admit only trivial $s_0^Z$ and use Table~\ref{tab:s2LUT} to decode $s_0^X$. However, even with this additional filtering, the \textit{estimated} $p_{\rm L}$, calculated using Eq.~\ref{eq:logical_error_rate}, continues to exhibit non-FT behavior, scaling linearly with $p$ even after post-selection.

To address this apparent discrepancy, we carefully analyze the errors contributing to the logical error rates, as detailed in App.~\ref{app:SteaneEC_pL}. We find that most of the recorded logical errors do not actually involve the $X_{\rm L} = X_1 X_4 X_7$ operator. When considering the \textit{true} logical error rate, Steane EC predictably outperforms FB EC (bold orange vs. bold blue curves in Fig.~\ref{fig:SteaneEC_vs_FBEC}a-b), even without post-selection. In addition, the \textit{true} logical error rates for both protocols exhibit the expected $p^2$ scaling in the low-noise regime. Remarkably, as shown in Fig.~\ref{fig:SteaneEC_vs_FBEC}b, incorporating $s_0^Z$-based post-selection reduces the number of \textit{true} logical errors per million shots for Steane EC to virtually zero within the examined noise range. This, along with an experimental demonstration using a fidelity-based metric~\cite{Postler_2024_SteaneEC}, confirms that Steane EC achieves the exceptionally low \textit{true} logical error rate expected by design. Our findings suggest that estimating the logical error rate solely based on the commutator with logical operators significantly overestimates the logical error rate for Steane EC, highlighting the need for a more accurate metric.
% As a result, in terms of practically measurable metrics, the \textit{Flag-Bridge} approach emerges as the more effective strategy

\section{Conclusion and Outlook}\label{Sec:conclusions}

In this work, we have explored various fault-tolerant approaches to prepare the $\ket{0}_{\rm L}$ state of the [[7,1,3]] code through numerical simulations. Our analysis specifically incorporated several practical considerations by: (i) proposing compact implementations of the code on a 2D grid topology with minimal overhead, (ii) ensuring error-correction readiness of the encoding circuits, and (iii) extending our analysis to include both noisy encoding and noisy error correction. In particular, we focused on a flag-based approach, which utilizes flag qubits to ensure fault tolerance and facilitate interactions between data and syndrome qubits on a topology with limited connectivity (\textit{Flag-Bridge} encoding). This approach supports both encoding and error-correction. We compared it against a single-ancilla, verification-based encoding method (\textit{GotoRL} encoding) and the Steane EC protocol, which achieves fault tolerance through transversal CNOT gates~\cite{Steane1996_FTQEC}.

We first demonstrated the advantage of the \textit{Flag-Bridge} approach when considering only the encoding stage. Assuming perfect error correction, \textit{Flag-Bridge} encoding achieves a 3.4\% encoding pseudo-threshold, more than double that of \textit{GotoRL} encoding, albeit with a lower acceptance rate. The inclusion of six flag qubits, as opposed to one in \textit{GotoRL} encoding, enables straightforward tuning of the performance-efficiency trade-off by expanding the set of acceptable flag outcomes. This flexibility accommodates various hardware platforms with differing limitations and requirements, without necessitating additional runs or modifications.

When accounting for noisy error-correction, the \textit{Flag-Bridge} approach supports an efficient \textit{hybrid} protocol that integrates encoding and error-correction using the same syndrome extraction circuits. Remarkably, this \textit{hybrid} \textit{Flag-Bridge} protocol maintains a performance advantage over \textit{GotoRL} encoding (0.13\% versus 0.06\% \textit{pseudo-threshold}) while matching the performance of a standalone noisy error-correction protocol with perfect encoding. Notably, the optimal performance is achievable with a sub-optimal decoding strategy that employs flag information only for post-selection. This observation suggests a straightforward abort-and-restart strategy whenever a flag is triggered, simplifying the implementation on control devices.

\begingroup
\setlength{\tabcolsep}{2.5pt}
\begin{table*}[t!]
    \begin{minipage}{0.48\textwidth}
    \begin{center}
    \begin{tabular}{|ccccccc|} 
        \multicolumn{7}{c}{(a) Flag pattern: 01 00 00} \\
        \hline Index & Gate & Fault & Initial Pauli & $s$ & $f$ & Error \rule{0pt}{2.5ex} \\
        \hline \hline
        2 & CX[8,9] & -$X$ & -\,-\,-\,-\,-\,-\,-$|$-\,-$X$- & 000 & 010000 & -\,-\,-$X$-\,-\,- \\
		2 & CX[8,9] & -$Y$ & -\,-\,-\,-\,-\,-\,-$|$-\,-$Y$- & 100 & 010000 & -\,-\,-$X$-\,-\,- \\
		2 & CX[8,9] & $Z$$X$ & -\,-\,-\,-\,-\,-\,-$|$-$Z$$X$- & 100 & 010000 & -\,-\,-$X$-\,-\,- \\
		2 & CX[8,9] & $Z$$Y$ & -\,-\,-\,-\,-\,-\,-$|$-$Z$$Y$- & 000 & 010000 & -\,-\,-$X$-\,-\,- \\
		6 & CX[9,3] & $X$- & -\,-\,-\,-\,-\,-\,-$|$-\,-$X$- & 000 & 010000 & -\,-\,-\,-\,-\,-\,- \\
		6 & CX[9,3] & $X$$X$ & -\,-\,-$X$-\,-\,-$|$-\,-$X$- & 000 & 010000 & -\,-\,-$X$-\,-\,- \\
		6 & CX[9,3] & $X$$Y$ & -\,-\,-$Y$-\,-\,-$|$-\,-$X$- & 001 & 010000 & -\,-\,-$Y$-\,-\,- \\
		6 & CX[9,3] & $X$$Z$ & -\,-\,-$Z$-\,-\,-$|$-\,-$X$- & 001 & 010000 & -\,-\,-$Z$-\,-\,- \\
		6 & CX[9,3] & $Y$- & -\,-\,-\,-\,-\,-\,-$|$-\,-$Y$- & 100 & 010000 & -\,-\,-\,-\,-\,-\,- \\
		6 & CX[9,3] & $Y$$X$ & -\,-\,-$X$-\,-\,-$|$-\,-$Y$- & 100 & 010000 & -\,-\,-$X$-\,-\,- \\
		6 & CX[9,3] & $Y$$Y$ & -\,-\,-$Y$-\,-\,-$|$-\,-$Y$- & 101 & 010000 & -\,-\,-$Y$-\,-\,- \\
		6 & CX[9,3] & $Y$$Z$ & -\,-\,-$Z$-\,-\,-$|$-\,-$Y$- & 101 & 010000 & -\,-\,-$Z$-\,-\,- \\
		7 & CX[8,9] & -$X$ & -\,-\,-\,-\,-\,-\,-$|$-\,-$X$- & 000 & 010000 & -\,-\,-\,-\,-\,-\,- \\
		7 & CX[8,9] & -$Y$ & -\,-\,-\,-\,-\,-\,-$|$-\,-$Y$- & 000 & 010000 & -\,-\,-\,-\,-\,-\,- \\
		7 & CX[8,9] & $Z$$X$ & -\,-\,-\,-\,-\,-\,-$|$-$Z$$X$- & 100 & 010000 & -\,-\,-\,-\,-\,-\,- \\
		7 & CX[8,9] & $Z$$Y$ & -\,-\,-\,-\,-\,-\,-$|$-$Z$$Y$- & 100 & 010000 & -\,-\,-\,-\,-\,-\,- \\
        \hline
    \end{tabular}
    % \caption{Pattern 010000: H[8], CX[8,7], CX[8,9], CX[8,1], CX[8,2], CX[7,0], CX[9,3], CX[8,9], CX[8,7], H[8], }
    \end{center}
    \end{minipage}
    \hfill
    \begin{minipage}{0.48\textwidth}
    \begin{center}
    \begin{tabular}{|ccccccc|} 
        \multicolumn{7}{c}{(b) Flag pattern: 11 00 00} \\
        \hline Index & Gate & Fault & Initial Pauli & $s$ & $f$ & Error \rule{0pt}{2.5ex} \\
        \hline \hline
		2 & CX[8,9] & $X$- & -\,-\,-\,-\,-\,-\,-$|$-$X$-\,- & 000 & 110000 & -$X$$X$-\,-\,-\,- \\
		2 & CX[8,9] & $X$$Z$ & -\,-\,-\,-\,-\,-\,-$|$-$X$$Z$- & 100 & 110000 & -$X$$X$-\,-\,-\,- \\
		2 & CX[8,9] & $Y$- & -\,-\,-\,-\,-\,-\,-$|$-$Y$-\,- & 100 & 110000 & -$X$$X$-\,-\,-\,- \\
		2 & CX[8,9] & $Y$$Z$ & -\,-\,-\,-\,-\,-\,-$|$-$Y$$Z$- & 000 & 110000 & -$X$$X$-\,-\,-\,- \\
		3 & CX[8,1] & $X$- & -\,-\,-\,-\,-\,-\,-$|$-$X$-\,- & 000 & 110000 & -\,-$X$-\,-\,-\,- \\
		3 & CX[8,1] & $X$$X$ & -$X$-\,-\,-\,-\,-$|$-$X$-\,- & 000 & 110000 & -$X$$X$-\,-\,-\,- \\
		3 & CX[8,1] & $X$$Y$ & -$Y$-\,-\,-\,-\,-$|$-$X$-\,- & 010 & 110000 & -$Y$$X$-\,-\,-\,- \\
		3 & CX[8,1] & $X$$Z$ & -$Z$-\,-\,-\,-\,-$|$-$X$-\,- & 010 & 110000 & -$Z$$X$-\,-\,-\,- \\
		3 & CX[8,1] & $Y$- & -\,-\,-\,-\,-\,-\,-$|$-$Y$-\,- & 100 & 110000 & -\,-$X$-\,-\,-\,- \\
		3 & CX[8,1] & $Y$$X$ & -$X$-\,-\,-\,-\,-$|$-$Y$-\,- & 100 & 110000 & -$X$$X$-\,-\,-\,- \\
		3 & CX[8,1] & $Y$$Y$ & -$Y$-\,-\,-\,-\,-$|$-$Y$-\,- & 110 & 110000 & -$Y$$X$-\,-\,-\,- \\
		3 & CX[8,1] & $Y$$Z$ & -$Z$-\,-\,-\,-\,-$|$-$Y$-\,- & 110 & 110000 & -$Z$$X$-\,-\,-\,- \\
		4 & CX[8,2] & $X$- & -\,-\,-\,-\,-\,-\,-$|$-$X$-\,- & 000 & 110000 & -\,-\,-\,-\,-\,-\,- \\
		4 & CX[8,2] & $X$$X$ & -\,-$X$-\,-\,-\,-$|$-$X$-\,- & 000 & 110000 & -\,-$X$-\,-\,-\,- \\
		4 & CX[8,2] & $X$$Y$ & -\,-$Y$-\,-\,-\,-$|$-$X$-\,- & 011 & 110000 & -\,-$Y$-\,-\,-\,- \\
		4 & CX[8,2] & $X$$Z$ & -\,-$Z$-\,-\,-\,-$|$-$X$-\,- & 011 & 110000 & -\,-$Z$-\,-\,-\,- \\
		4 & CX[8,2] & $Y$- & -\,-\,-\,-\,-\,-\,-$|$-$Y$-\,- & 100 & 110000 & -\,-\,-\,-\,-\,-\,- \\
		4 & CX[8,2] & $Y$$X$ & -\,-$X$-\,-\,-\,-$|$-$Y$-\,- & 100 & 110000 & -\,-$X$-\,-\,-\,- \\
		4 & CX[8,2] & $Y$$Y$ & -\,-$Y$-\,-\,-\,-$|$-$Y$-\,- & 111 & 110000 & -\,-$Y$-\,-\,-\,- \\
		4 & CX[8,2] & $Y$$Z$ & -\,-$Z$-\,-\,-\,-$|$-$Y$-\,- & 111 & 110000 & -\,-$Z$-\,-\,-\,- \\
		7 & CX[8,9] & $X$$X$ & -\,-\,-\,-\,-\,-\,-$|$-$X$$X$- & 000 & 110000 & -\,-\,-\,-\,-\,-\,- \\
		7 & CX[8,9] & $X$$Y$ & -\,-\,-\,-\,-\,-\,-$|$-$X$$Y$- & 000 & 110000 & -\,-\,-\,-\,-\,-\,- \\
		7 & CX[8,9] & $Y$$X$ & -\,-\,-\,-\,-\,-\,-$|$-$Y$$X$- & 100 & 110000 & -\,-\,-\,-\,-\,-\,- \\
		7 & CX[8,9] & $Y$$Y$ & -\,-\,-\,-\,-\,-\,-$|$-$Y$$Y$- & 100 & 110000 & -\,-\,-\,-\,-\,-\,- \\
        \hline
    \end{tabular}
    % \caption{Pattern 110000: H[8], CX[8,7], CX[8,9], CX[8,1], CX[8,2], CX[7,0], CX[9,3], CX[8,9], CX[8,7], H[8], }
    % \label{tab:flagpattern_110000}
    \end{center}
    \end{minipage}
\caption{Summary of faults for two exemplary flag patterns in \textit{Flag-Bridge} encoding. The `01 00 00' flag pattern results in final errors on the data qubits of at most weight-1, allowing it to be immediately included in the accepted flag set. In contrast, the `11 00 00' flag pattern, which may lead to higher-weight errors, requires an additional correction, such as $X_3$, to ensure that all final errors from a single fault are at most weight-1. The associated gate sequence for both cases is: H[8], CX[8,7], CX[8,9], CX[8,1], CX[8,2], CX[7,0], CX[9,3], CX[8,9], CX[8,7], H[8].}
\label{tab:flagpatterns}
\end{table*}
\endgroup

Compared to Steane EC, the \textit{Flag-Bridge} approach achieves better \textit{estimated} logical error rates, calculated using the commutator with the $Z_{\rm L}$ operator. This advantage stems from the additional flag qubits, which restrict error propagation to fewer initial faults. However, Steane EC outperforms in terms of \textit{true} logical error counts, which consider only errors containing the $X_{\rm L}$ operator, thereby justifying its use of twice as many ancilla qubits. Therefore, to fairly assess the experimental performance of Steane EC, a more accurate metric, such as a fidelity-based approach~\cite{Postler_2024_SteaneEC}, is essential.

% \nam[]{}

Our correction-ready encoding circuits are particularly well-suited for experimental implementation on platforms with current 2D nearest-neighbor connectivity, such as superconducting qubits. They are also compatible with architectures like trapped ions and neutral atoms, where reducing qubit movement can help mitigate errors~\cite{Ruster2014,Moses2023_racetrack, ovide2024scalingassigningresourcesion, loschnauer2024scalablehighfidelityallelectroniccontrol, Bluvstein2024}. Although we have focused on post-selection, FT can also be ensured with active correction based on the flag information from the two-flag parity-check circuits that we have employed~\cite{delfosse2020shortshorstylesyndromesequences,reichardt2024demonstrationquantumcomputationerror}. Future work could extend this approach by adapting parallel syndrome extraction circuits to similarly compact layouts, further optimizing performance and scalability. Additionally, exploring the implementation of logical computation within these compact designs, whether through non-local entangling gates or measurement-based lattice surgery, represents a promising direction. As experimental platforms continue to advance and become more widely accessible, detailed theoretical and numerical studies tailored to the specific constraints of these platforms will be essential for enabling small-scale demonstrations and paving the way for the broader realization of quantum error correction.

\section*{Acknowledgments}

%This material is based upon work supported by the U.S. Department of Energy, Office of Science, National Quantum Information Science Research Centers, Quantum Systems Accelerator (HNN, KBW).   
This work was supported by the U.S. National Science Foundation under the Convergence Accelerator Program, Grant No. OIA-2134345, and by the Defense Advanced Research Projects Agency under Grant Number HR0011-24-9-0358.
HNN thanks John Paul Marceaux and Zack Weinstein for helpful conversation on various subjects pertaining to this work.

% \begin{figure*}[ht!]
%   \centering \includegraphics[width=0.9\textwidth]{figs/verify_circuit3.png}
%   \caption{Encoding performance comparison under circuit-level noise.}
% \label{fig:verifyFT}
% \end{figure*}

\appendix

\section{Verifying fault-tolerance} \label{app:verifyFT}
We verify the fault-tolerance of all circuits used in this work through brute-force Clifford simulations. Specifically, we manually inject a single Pauli fault for each possible error in the noise model described in Sec.~\ref{sec:numerical}, propagate the fault through the circuit, and record the resulting error $E$ on the data qubits. We also track whether syndrome and flag measurements indicate a change. Subsequently, we compute the lowest-weight stabilizer-equivalent form, $E_{\rm equiv}$, of the final error. A circuit is deemed fault-tolerant if every fault that results in a harmful error, defined as $\text{weight}(E_{\rm equiv})>1$, causes at least one flag to be triggered. As an example, Table~\ref{tab:verifyFT_circuit3} shows all harmful errors resulting from single faults in Circuit 3 of Fig.~\ref{fig:syndrome_extraction_circuits}b. In all cases, the flag measurements are non-trivial, thereby confirming that the circuit is indeed fault-tolerant.

% \begin{lstlisting}
% 1||00: 
% idx  gate   i1  i2  fault   init_pauli   final_pauli   synds  flags     E     E_equiv  wt  Zremoved  wt 
%  0   Init   0   -1    Y    Y------|----  Y------|----    1      00   Y------  Y------  1   X------   1  
%  0   Init   0   -1    Z    Z------|----  Z------|----    1      00   Z------  Z------  1   -------   0  
%  1   Init   1   -1    Y    -Y-----|----  -Y-----|----    1      00   -Y-----  -Y-----  1   -X-----   1  
%  1   Init   1   -1    Z    -Z-----|----  -Z-----|----    1      00   -Z-----  -Z-----  1   -------   0  
%  2   Init   2   -1    Y    --Y----|----  --Y----|----    1      00   --Y----  --Y----  1   --X----   1  
% \end{lstlisting}

\section{Handling non-trivial flag patterns}\label{app:nontrivial_flag_patterns}

Table~\ref{tab:flags_breakdown} provides a summary of flag outcomes that can be included during post-selection for balancing the trade-off between performance and efficiency in \textit{Flag-Bridge} encoding circuit. Our analysis categorizes these outcomes into two distinct types based on the brute-force simulation results detailed in App.~\ref{app:verifyFT}. Sorting the outcomes by flag patterns, we examine the final errors on the data qubits to identify the appropriate handling for each pattern.

The first type, such as `01 00 00', corresponds to errors of at most weight-1, as shown in Table~\ref{tab:flagpatterns}a, similar to the trivial pattern. These patterns naturally satisfy the FT condition and can be included without any further modification. The second type, marked with * in Table~\ref{tab:flags_breakdown}, such as `11 00 00', may lead to higher-weight errors. However, as seen in Table~\ref{tab:flagpatterns}b, these patterns can be made FT-compatible through simple corrections. For example, applying an $X_3$ correction when the `11 00 00' pattern is observed reduces the two-qubit $X$ error to a single-qubit error while introducing, at most, a weight-1 error in previously error-free cases. By systematically identifying and addressing each pattern, we can leverage the additional information from flag measurements to tailor \textit{Flag-Bridge} encoding circuit to our need.

% \begin{figure*}[t!]
%   \centering \includegraphics[width=0.8\textwidth]{figs/flagpattern_010000.png}
%   \caption{Encoding performance comparison under circuit-level noise.}
% \label{tab:flagpattern_010000}
% \end{figure*}

% \begin{figure*}[t!]
%   \centering \includegraphics[width=0.8\textwidth]{figs/flagpattern_110000.png}
%   \caption{Encoding performance comparison under circuit-level noise.}
% \label{tab:flagpattern_110000}
% \end{figure*}

\section{FB EC - $f_1s_2$ LUT} \label{app:FBEC_f1s2LUT}

In the analysis of FB EC in Sec.~\ref{subsubsec:FBEC_analysis}, we introduce the use of an additional lookup table that incorporates flag information from the first round ($f_1$) to decode the syndrome from the second round ($s_2$), instead of exclusively post-selecting on trivial flag outcomes. To construct this LUT, we begin by performing the Clifford simulation detailed in App.~\ref{app:verifyFT} for each stabilizer, generating $f_1$. Subsequently, we simulate a perfect round of error correction to obtain $s_2$. For each distinct combination of $f_1$ and $s_2$, we verify that the resulting final errors on the data qubits are unique (up to stabilizer equivalence). This unique mapping establishes the desired LUT. Furthermore, since all $Z$ errors on the $\ket{0}_{\rm L}$ state reduce to at most single-qubit errors, we can use the standard LUT in Table~\ref{tab:s2LUT} to decode $Z$ errors. As a result, the additional LUT is only required for decoding $X$ errors. Specifically, it uses the $X$-part of $f_1$ and $Z$-part of $s_2$, as summarized in Table~\ref{tab:f1Xs2ZLUT}. 

\begingroup
\setlength{\tabcolsep}{5pt}
\begin{table}[t!]
\begin{center}
    \begin{tabular}{|c|c|c|} 
        \hline $f_1^X$ & $s_2^Z$ & Error (Recovery) \rule{0pt}{2.5ex} \\
        \hline \hline
		10 00 00 & 100 & $X_1$ \\
		01 00 00 & 101 & $X_4$ \\
		11 00 00 & 001 & $X_5X_6$ \\
		11 00 00 & 111 & $X_3$ \\
		11 00 00 & 000 & $I$ \\
		10 00 00 & 000 & $I$ \\
		01 00 00 & 000 & $I$ \\
		00 10 00 & 001 & $X_5X_6$ \\
		00 01 00 & 011 & $X_6$ \\
		00 11 00 & 010 & $X_5$ \\
		00 10 00 & 111 & $X_3$ \\
		00 10 00 & 000 & $I$ \\
		00 11 00 & 000 & $I$ \\
		00 01 00 & 000 & $I$ \\
		00 00 01 & 111 & $X_3$ \\
		00 00 10 & 010 & $X_6X_7$ \\
		00 00 11 & 101 & $X_4$ \\
		00 00 10 & 011 & $X_6$ \\
		00 00 10 & 000 & $I$ \\
		00 00 11 & 000 & $I$ \\
		00 00 01 & 000 & $I$ \\
        \hline
    \end{tabular}
    \caption{Augmented look-up-table for the FT EC protocol of [[7,1,3]] Steane code. This LUT utilizes flag outcomes from the first round ($f_1^X$) and syndrome outcomes from the second round ($s_2^Z$) to determine the required recovery operation for correcting $X$ errors.}
    \label{tab:f1Xs2ZLUT}
\end{center}
\end{table}
\endgroup

\begin{figure*}[t!]
  \centering \includegraphics[width=1\textwidth]{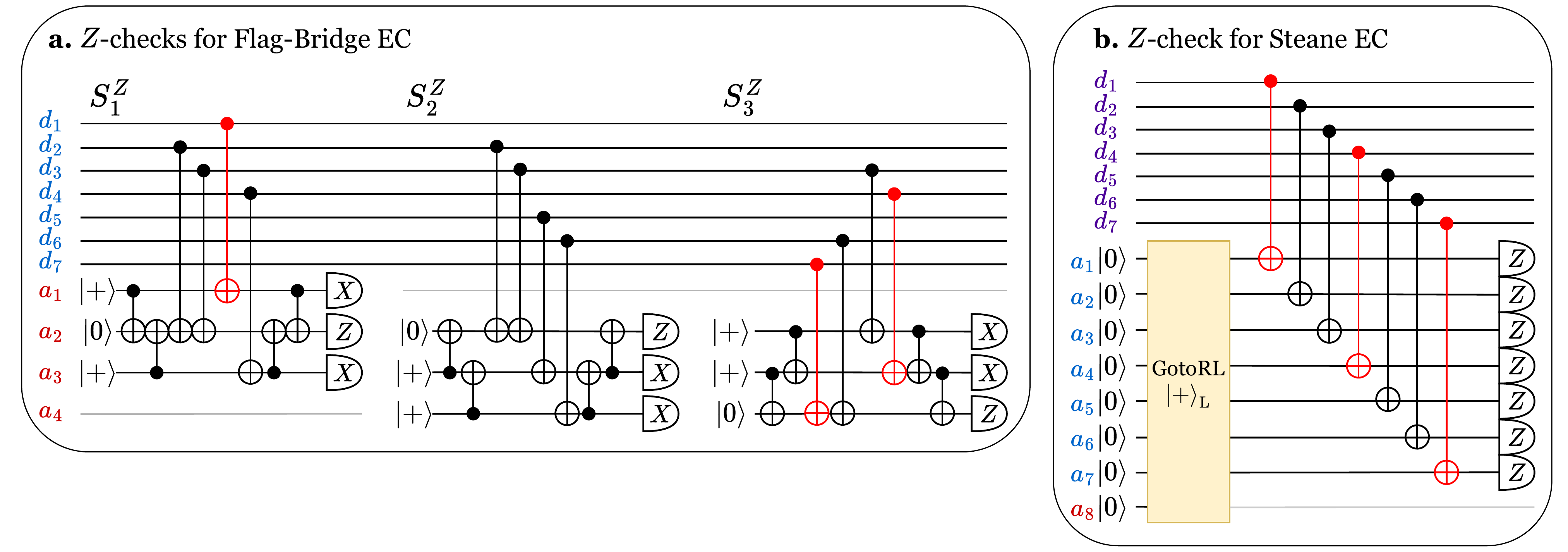}
  \caption{\textbf{Fault locations contributing to the \textit{estimated} logical error rate $p_{\rm L}$.} Gates highlighted in red denote fault locations that can induce single-qubit $X$ errors on qubits 1, 4, and 7, contributing to $p_{\rm L}$ as estimated via the commutation relation with $Z_{\rm L}=Z_1Z_4Z_7$. Earlier $X$ errors are excluded from this analysis, as they would trigger non-trivial $Z$-type syndrome outcomes $s_0^Z$. a) In the FB EC protocol, the potentially harmful fault locations are the final gates interacting with qubits 1, 4, and 7. b) In the $Z$-check circuit of Steane EC, three analogous fault locations similarly contribute to $p_{\rm L}$ estimation.}
\label{fig:fault_comparison}
\end{figure*}

\section{Steane EC - Logical error rate analysis}\label{app:SteaneEC_pL}

In Sec.~\ref{subsubsec:SteaneEC_analysis}, we observe that in the low noise regime, the logical error rate $p_{\rm L}$ for Steane EC, even after $s_0^Z$-post-selection, is worse than that of FB EC, despite relying only on transversal CNOT gates. We investigate this discrepancy by analyzing the calculation of \textit{estimated} $p_{\rm L}$ based on the commutation relation with $Z_{\rm L}=Z_1Z_4Z_7$, which requires accounting for all possible $X$ and $Y$ faults on qubits 1, 4, and 7. 

We focus on the \textit{hybrid} protocols for both \textit{Flag-Bridge} and Steane EC, where the $X$-part is executed as both the encoding and parity-check stage, followed by the $Z$-part. Faults arising during the $X$-part are detected by the subsequent $Z$-checks, resulting in non-trivial syndrome $s_0^Z$. Thus, by restricting our analysis to shots with trivial $s_0^Z$, for both FB EC and Steane EC, we can ignore $X$ and $Y$ faults from the $X$-part and focus on those originating from the $Z$-part.

Fig.~\ref{fig:fault_comparison} illustrates the $Z$-checks for both \textit{Flag-Bridge} and Steane EC protocols, with red highlighting gates that may introduce $X$ and $Y$ errors on qubits 1, 4, and 7. Since $X$ errors propagate from control to target during CNOT gates, only gates directly connected to qubits 1, 4, and 7 are potentially harmful. In the $Z$-part of FB EC protocol, two CNOT gates are connected to qubit 4 - one in $S_1^Z$ and the other in $S_3^Z$. If the first CNOT gate induces an $X$ error on qubit 4, the second CNOT triggers the last syndrome, producing a non-trivial $s_0^Z$. Therefore, only the final instances of gates connected to qubits 1, 4, and 7 are relevant.

Although the \textit{Flag-Bridge} and Steane EC protocols initially appear to have the same number of potentially harmful gates contributing to the logical error rate, the actual error channels differ. For FB EC, only $XI$ fault is permissible, as $XX$ and $XY$ will trigger the syndrome, and $XZ$ will trigger a flag. In contrast, the Steane EC circuit, lacking flag qubits, allows both $XI$ and $XZ$ faults, effectively doubling the contribution to the logical error rate. Indeed, for lower physical error rates in Fig.~\ref{fig:SteaneEC_vs_FBEC}b, the logical error count for Steane EC is approximately double that of FB EC. These findings further support that estimating the logical error rate based on commutation relation can significantly overestimate its value for Steane EC.

% \section{Additional Data}

% \begin{figure*}[ht!]
%   \centering \includegraphics[width=\textwidth]{figs/encoding_1QECcycle_all.pdf}
%   \caption{Encoding + 1 QEC cycle performance comparison under circuit-level noise.}
% \label{fig:encodinS_1QECcycle_all.pdf}
% \end{figure*}

\bibliography{citation}% Produces the bibliography via BibTeX.

\end{document}